\documentclass[aps,prd,reprint,superscriptaddress]{revtex4-2}

\usepackage{comment}
\usepackage[utf8]{inputenc} 
\usepackage{graphicx,color,overpic,mathtools}
\usepackage{amsthm,amsmath,amssymb,hyperref,mathrsfs}
\usepackage{amsfonts}
\usepackage{braket,bm,bbm,setspace}
\PassOptionsToPackage{normalem}{ulem}
\usepackage{ulem} 
\usepackage{physics}
\usepackage{float}
\usepackage[makeroom]{cancel}
\usepackage[english]{babel}
\usepackage{graphicx}
\usepackage{soul}
\graphicspath{ {images/} }
\addto\captionsspanish{}
\hypersetup{
    colorlinks=false,
    pdfborder={0 0 0},
} 
\definecolor{emerald}{rgb}{0.31, 0.78, 0.47}

\begin{document} 

\title{Emergent gauge symmetries: Yang-Mills theory}

\author{Carlos Barceló}
\email{carlos@iaa.es}
\affiliation{Instituto de Astrofísica de Andalucía (IAA-CSIC), Glorieta de la Astronomía, 18008 Granada, Spain}
\author{Raúl Carballo-Rubio}
\email{raul.carballorubio@ucf.edu}
\affiliation{Florida Space Institute, University of Central Florida, 12354 Research Parkway, Partnership 1, Orlando, Florida 32826, USA}
\author{Luis J. Garay}
\email{luisj.garay@ucm.es}
\affiliation{Departamento de Física Teórica and IPARCOS, Universidad Complutense de Madrid, 28040 Madrid, Spain}
\affiliation{Instituto de Estructura de la Materia (IEM-CSIC), Serrano 121, 28006 Madrid, Spain}
\author{Gerardo García-Moreno}
\email{gerargar@ucm.es}
\affiliation{Departamento de Física Teórica and IPARCOS, Universidad Complutense de Madrid, 28040 Madrid, Spain}
\affiliation{Instituto de Astrofísica de Andalucía (IAA-CSIC), Glorieta de la Astronomía, 18008 Granada, Spain}

\begin{abstract}{Gauge symmetries remove unphysical states and guarantee that field theories are free from the pathologies associated with these states. In this work we find a set of general conditions that guarantee the removal of unphysical states in field theories describing interacting vector fields. These conditions are obtained through the extension of a mechanism for the emergence of gauge symmetries proposed in a previous article [C. Barceló \emph{et al.} JHEP 10 (2016) 084] in order to account for non-Abelian gauge symmetries, and are the following: low-energy Lorentz invariance, emergence of massless vector fields describable by an action quadratic in those fields and their derivatives, and self-coupling to a conserved current associated with specific rigid symmetries. Using a bootstrapping procedure, we prove that these conditions are equivalent to the emergence of gauge symmetries and, therefore, guarantee that any theory satisfying them must be equivalent to a Yang-Mills theory at low energies.}
\end{abstract}

\maketitle

\section{Introduction}

The search for a theory of quantum gravity, i.e., a theory which combines the principles of general relativity and quantum mechanics, has been one of the key cornerstones in fundamental physics of the last century. Until now, it has not been possible to find a completely satisfactory theory, although there are  many illuminating approaches \cite{carlip2001}. Among these approaches toward constructing a theory of quantum gravity, we could distinguish whether the geometrical and spacetime notions, characteristic of general relativity, are emergent or not. On the one hand, we have the theories that consider the geometrical degrees of freedom and the dynamics of spacetime as fundamental. Such theories typically try to apply quantization schemes to these degrees of freedom seeking for a background independent theory of quantum gravity. Canonical quantum gravity in its modern formulation in terms of loop quantization \cite{thiemann2007} is the most popular approach within this category. On the other hand, we have approaches in which the fundamental degrees of freedom are not taken to be the spacetime itself but such a concept emerges with all of the properties of general relativity in some regimes of the theory, typically in low-energy limits. Within this category, we would include string theory~\cite{polchinski1998a,Polchinski1998b}, but also theories that start from condensed-matter-like systems as the substratum for emergence~\cite{Barcelo2005}. One of the main challenges of this last approach is to explain when and how a diffeomorphism gauge symmetry can emerge in physical systems that do not include it in their microscopic description \cite{carlip2013}. A related problem is the possible emergence of gauge symmetries in particle physics \cite{Witten2017}. The emergent paradigm is, somehow, the opposite direction to the one that has been explored the most by the community, which is enlarging the gauge symmetry group of the Standard Model at high energies, instead of breaking it \cite{Bass2020}. Grand Unified Theories~\cite{raby2017} or Technicolor~\cite{lane1993} are archetypal examples of this direction of work.

In this work we pursue a program for understanding emergent gauge symmetries in general, introduced by some of the authors in a previous work \cite{barcelo2016}. In that work we presented a linear system closely related to electrodynamics and described how an effective gauge symmetry naturally emerges, with the only prerequisite of having a Lorentz covariant description with massless fields. Before addressing the more convoluted problem of diffeomorphism emergence, in this work we shall analyze how the Abelian description in~\cite{barcelo2016} generalizes to non-Abelian symmetries, and then, also interacting nonlinear theories of relativistic fields. Such a theory offers new difficulties that are not present in the linear theory, and would bridge the gap between the simple linear electromagnetic case and the more complicated gravitational case, that we leave for a future work. 

We begin by reviewing in abstract terms (Sec. \ref{section_2}) the main ideas of the mechanism presented in \cite{barcelo2016} for the emergence of gauge symmetries. We emphasize the differences between physical and gauge symmetries. In \cite{barcelo2016} it was proved that these abstract ideas
are clearly implemented in a linear model tightly related to electrodynamics. In Sec. \ref{section_3} we will introduce an equivalent linear system but now with a collection of Lorentzian fields $A^a_\mu$ which in a second stage will be subject to interactions among them. At this stage we again find an emergence of gauge symmetries. We approach the problem by studying the most general Lorentz-invariant quadratic action for a set of relativistic fields and see how the gauge symmetry of a system of uncoupled Maxwell systems emerges.

The more novel issues of the present work start when discussing how to deform the linear theory into a nonlinear theory (Sec. \ref{section4}). To generate a nonlinear theory we apply a bootstrapping mechanism. The bootstrapping of theories of Yang-Mills type was worked out already in the seminal work by Deser \cite{deser1970}, with the starting point being a linear theory of vector fields invariant under both Lorentz and gauge transformations. Here we drop the assumption of gauge invariance for the linear theory, which is essential in order to discuss the emergence of gauge symmetries following the ideas discussed in \cite{barcelo2016}. Hence, our bootstrapping procedure represents an extension of Deser's analysis. In fact, we show that the results in the latter work carry out to this extended framework and are therefore more general, in the sense that they still apply even when the assumption of gauge invariance of the linear theory is relaxed. We will explain that the reason behind the result is the existence of consistency conditions necessary for the bootstrapping procedure to work and that turn out to be equivalent to the nonlinear theory being gauge invariant. We will discuss how these consistency conditions lead to the construction of a family of theories with emergent gauge symmetries, characterized by the choice of a specific Lie algebra of the same dimension as the number of relativistic fields involved in the construction. This emergent gauge symmetry guarantees that the theory is free of ghosts or any other sort of classical instability (in other words, the bootstrapping procedure implies the decoupling of unphysical states that would give rise to pathologies). Our analysis will also illustrate some aspects that were not addressed explicitly in previous works, such as the role played by boundary terms in the iterative procedure and their interplay with the possible uniqueness of the latter. We will also discuss the bootstrapping procedure for the charged matter sector, and discuss related issues such as the physical interpretation in our formalism of the so-called Gribov copies.

Weinberg-Witten theorems \cite{weinberg1980} and Marolf's theorem \cite{marolf2015} are often invoked as impediments toward having a successful framework in which gauge symmetries of Yang-Mills and gravitational character can emerge. In fact, these results are often used to state that such a program is condemned to fail from the beginning. One of the aims of this work is to provide a concrete framework   which  illustrates how these impediments can be bypassed, while also allowing to describe the emergence of gauge symmetries.

\textit{Notation and conventions.} We work in four dimensions, using Minkowskian coordinates and the signature $(-,+,+,+)$. The symbol $\partial_{\mu}$ represents the ordinary derivative in Minkowski spacetime. We use greek $(\mu, \nu...)$ indices for the spacetime indices, latin indices from the beginning of the alphabet $(a,b...)$ for the internal indices on the space of gauge fields, and from the middle of the alphabet $(i,j...)$ for the internal indices within the flavor space of matter fields.

\section{Emergence of gauge symmetries}
\label{section_2}

In this section, we   first clarify the meaning of gauge symmetries, making special emphasis on the fact that local symmetries, i.e., those whose generators are functions of the position in spacetime, are not necessarily gauge symmetries. We   then review the mechanism presented in \cite{barcelo2016} for the emergence of gauge symmetries in certain systems.

\subsection{What are gauge symmetries?}

For finite-dimensional systems it is straightforward to discern whether a given symmetry is physical or gauge by looking at the group parameters; physical symmetries have a finite set of parameters, while gauge symmetries are always parametrized by functions \cite{banados2016,henneaux1992}. In the infinite-dimensional case of a field theory, this shortcut does no longer work because all symmetries are now parametrized by functions \cite{banados2016}. 

To circumvent this problem, one can perform a canonical analysis in phase space following the procedure introduced by Dirac \cite{dirac1964}, in which the existence of gauge symmetries will manifest in the appearance of first-class constraints \cite{wipf1993}. Alternatively, one can study the Noether currents associated with symmetries, given that gauge symmetries are characterized by having identically zero Noether charges, contrary to physical symmetries whose charges are nontrivial and can be used as coordinates parametrizing the space of solutions of the theory. Throughout this work we will follow the second procedure, although both are equivalent. 

More specifically, the current associated with a general gauge symmetry can be written as

\begin{equation}
J^{\mu} = W^{\mu} + S^{\mu},
\end{equation}
where $W^{\mu}$ is zero on-shell ($W^{\mu} \mid_{\mathcal{S}} = 0$, being $\mathcal{S}$ the space of solutions of the theory) and $S^{\mu} = \partial_{\nu} N^{[\nu \mu]}$ is a superpotential, i.e., the divergence of an antisymmetric tensor $N^{\mu \nu}$, that is identically (i.e., not only on-shell) conserved. Once the charge is computed on-shell as an integral of $J^0$, it is clear that the first term does not contribute. Furthermore, the second term always produces a boundary term evaluated at the spatial boundary of the spacetime in virtue of Gauss theorem. Suitable boundary conditions supplementing the equations of motion typically guarantee fall-off conditions such that this contribution also vanishes, rendering a trivial Noether charge \cite{julia1998}.

\subsection{A mechanism for the emergence of gauge symmetries from physical ones}

The mechanism presented in \cite{barcelo2016} for the emergence of gauge symmetries is based on this characterization of gauge symmetries. In a system displaying no gauge symmetries \textit{a priori}, all Noether currents are nontrivial. However, if certain conditions are met that turn out to make some of these Noether currents trivial, the corresponding symmetries could be considered as emergent gauge symmetries.

Let us consider a field theory depending on a collection of fields $\{ \phi_a \}_{a \in J}$. Let us refer to them collectively as $\Phi$. For simplicity we require that the system is free of gauge symmetries\footnote{It is equally possible to consider the emergence of additional gauge symmetries in a system in which some of the symmetries are already gauge; however, this makes the discussion more convoluted without providing additional conceptual insights.}. This means that all of the symmetries the theory might display will have nonvanishing Noether charges. These charges can be used to parametrize the space of solutions to the equations of motion $\mathcal{S}$. This follows from the fact that there always exists a complete set of symmetries whose Noether charges parametrize the space of initial conditions of the dynamical equations of the theory \cite{aldaya2016}.

These are in general complicated, contact and nonpoint symmetries \cite{aldaya2016} that cannot  be generally found explicitly, although their existence is always guaranteed as long as the initial value problem is well posed. Otherwise, we would have that every system can be explicitly solved in terms of these charges, which is not the case. Integrable systems are those for which these symmetries can be explicitly found out.

Let us now introduce a set of constraints defined as $\Psi=0$. This set of constraints, that might be satisfied only approximately, can be understood as the decoupling of some of the degrees of freedom. The situations we are interested in are those in which the subspace $\mathcal{U} \subset \mathcal{S}$ which we define as the subspace of $\mathcal{S}$ for which $\Psi = 0$ is nontrivial, in the sense that there are nontrivial solutions $\Phi \neq 0$ for which $\Psi = 0$. Requiring the subspace $\mathcal{U}$ to be nontrivial is a necessary condition that determines whether a choice of constraints $\Psi=0$ is suitable for our aim.

In terms of Noether charges, we can define $\mathcal{Q}$ as the complete set of charges that parametrize $\mathcal{S}$. One can always find a parametrization such that the condition $\Psi = 0$ amounts to the selection of a subset $\mathcal{Q}_{\Psi}^\perp \subset \mathcal{Q}$ with the requirement $\mathcal{Q}_{\Psi}^\perp=0$. There exist other sets of charges $\mathcal{Q}_{\Psi}^{\not\perp}$ which parametrize the different solutions within the set $\mathcal{U}$, i.e., there are different systems of Noether-charge coordinates that one can use to distinguish solutions inside $\mathcal{U}$. In principle, one can decide to use as internal Noether coordinates in $\mathcal{U}$, those associated with symmetries that can be defined within $\mathcal{U}$, i.e., symmetries which leave the condition $\Psi=0$ untouched. Let us denote by $\mathcal{Q}_{\Psi}^\parallel$ those coordinates. Then, depending of the specific system and the condition $\Psi$, we can have two different scenarios.

\textbf{Nonemergence of gauge symmetry:} One can find that the set of Noether charges associated with symmetries $\mathcal{Q}_{\Psi}^\parallel$ that preserve the subspace $\mathcal{U}$ (by assumption there will always be some of them) is a proper system of Noether coordinates in $\mathcal{U}$. This is the standard situation one can find. These charges essentially parametrize $\mathcal{U}$ without redundancies. Then, the projection onto $\mathcal{U}$ solely removes the freedom associated with the value of $\Psi$, or, equivalently it only leaves the freedom parametrized by the Noether charges $\mathcal{Q}_{\Psi}^{\parallel}$.

\textbf{Emergence of gauge symmetry:} This happens when some of the physical symmetries preserving the subset $\mathcal{U}$ have trivial Noether charges when restricted to  $\mathcal{U}$, i.e. when $\Psi=0$. In this case, we cannot find a properly constructed internal Noether coordinate system in $\mathcal{U}$. Not all the different physical solutions in $\mathcal{U}$ can be distinguished by using only operations within $\mathcal{U}$. However, they are clearly distinguishable from the point of view of the entire theory. If all the probes one has about the system were through these charges, one could conclude that there exist equivalent classes of solutions in $\mathcal{U}$ which, being indistinguishable, can be understood as representing a single physical configuration. These equivalence classes correspond to the emergent gauge orbits that appear within this subspace. Then, apart from the reduction of freedom intrinsic to the projection onto $\mathcal{U}$, in practice there is an additional reduction of freedom since some of the configurations within $\mathcal{U}$ are physically identified (they belong to the same equivalence class that we have introduced). All in all, this process can be interpreted as the entire elimination of dynamical degrees of freedom (meaning whole solutions; recall that each degree of freedom is defined by a pair of initial conditions) when looking only at the  sector of the theory characterized by $\mathcal{U}$. 

This mechanism for the emergence of gauge symmetries strongly relies on the naturalness with which a specific system of effective equations and constraints $\Psi=0$  might appear in a low-energy regime of a possibly much more complicated theory. Although quite abstract, as here formulated, this mechanism for the emergence of gauge symmetries has been proved to work for an extension of electrodynamics in \cite{barcelo2016}. We will pursue here a generalization of the mechanism for Yang-Mills theory, the main novelty being the nonlinear nature of the latter theory. The existence of a suitable bootstrapping procedure connecting nonlinear theories with their linear limit will prove crucial for the definition of the decoupling conditions in the nonlinear theory.

\section{Linearized Yang-Mills theory: the emergence paradigm}
\label{section_3}

Let us begin with the emergence of the linearized Yang-Mills gauge symmetry. The starting point of our discussion will be the most general Lorentz invariant action quadratic in a collection of vector fields $A^{a}_{\mu}$ and their first-order derivatives, where the latin indices ($a,b,c...$) run from $1$ to $N$. At this level, these indices are just labels without a deeper physical meaning. Modulo boundary terms which do not modify the equations of motion, the most general action that we can construct for such a theory is the following: 
\begin{align}
S = \int d^4x \bigg [ &- P_{ab} \frac{1}{4}F^{a}_{ \mu \nu} F^{b \mu \nu}  + \frac{1}{2} \xi_{ab} (\partial_{\mu} A^{a \mu} ) (\partial_{\nu} A^{b \nu}) \nonumber\\ 
&-  \frac{1}{2} M_{ab}A^{a}_{\mu} A^{b \mu} + A_{a \mu} j^{a \mu} \bigg], 
\end{align}
where the tensor $F^{a}_{ \mu \nu}$ is defined as 
\begin{equation}\label{eq:lsdef}
F^{a}_{\mu \nu} = 2 \partial_{[\mu} A^{a}_{\nu]};
\end{equation} 
$P_{ab}, \xi_{ab}, M_{ab}$ are symmetric, constant matrices; and $j^{a \mu} $ are conserved currents \mbox{$\partial_{\mu} j^{a \mu} = 0$} representing the matter field content. Furthermore, we require that the matrix $P_{ab}$ be nondegenerate and  positive definite, as otherwise some of the equations of motion will not be of second order and the system will not correspond to $4N$ local propagating degrees of freedom with the appropriate sign for the kinetic term.  We can eliminate the matrix $P_{ab}$ at the expense of changing the matrices $\xi_{ab}$ and $M_{ab}$. Since $P_{ab}$ is a real   symmetric matrix, we can always find an invertible matrix $R^{a}_{b}$ that transforms it to the identity, i.e. that $R^{a}_{b} R^{c}_{d} P_{ac} = \delta_{bd}$. Then the field transformation $A^{a}_{\mu} \rightarrow R^{a}_{b} A^{b}_{\mu}$   (which also changes the matrices $\xi_{ab},M_{ab}$, and the current $j^{a \mu}$, although we will keep the same symbols to avoid a more cumbersome notation), provides the following general action:
\begin{align}\label{eq:zeroact}
S_0 = \int d^4x \bigg [ &-\frac{1}{4}F^{a}_{ \mu \nu} F^{b \mu \nu} + \frac{1}{2} \xi_{ab} (\partial_{\mu} A^{a \mu} ) (\partial_{\nu} A^{b \nu} ) \nonumber\\  &- \frac{1}{2} M_{ab}A^{a}_{\mu} A^{b \mu} + A_{a \mu} j^{a \mu} \bigg].
\end{align}
The Euler-Lagrange equations derived from this action are 
\begin{equation}
\partial_{\mu} F_{b}^{\mu \nu} - \xi_{ab} \partial^{\nu} \partial_{\mu} A^{a \mu} - M_{ab} A^{a \nu} = j_{b}^{\nu}.
\label{linearYM}
\end{equation}

The case $\xi_{ab} = M_{ab} = 0$  corresponds to the linearization of a Yang-Mills theory whose gauge group is of dimension $N$.  This linearization is equivalent to a system of decoupled Maxwell equations.

Let us focus on the general case in which both matrices $\xi_{ab}$ and $M_{ab}$ are nondegenerate. Since the currents $j^ a_\mu$ are  conserved, there is a physical symmetry of the theory, given by the following transformations: 
\begin{equation}
A_{\mu}^a \rightarrow A_{\mu} ^a + \partial_{\mu} \chi^a, \qquad j^{a \mu} \rightarrow j^{a \mu},
\label{linearizedgauge}
\end{equation}
where $\chi^a$ are not arbitrary functions as in the linearized Yang-Mills case, but they need to obey
\begin{equation}
\left(\xi_{ab} \Box + M_{ab} \right) \chi^{a} = 0.
\label{condition_gauge}
\end{equation}
Clearly, we need to impose boundary conditions such that $\chi^a$ vanish at infinity, ensuring that there are no zero modes. We can compute the conserved quantities associated with such symmetries by considering the previous transformation to be infinitesimal and applying Noether's theorem. If we compute the current $J^{\mu}_{\chi}$ associated with the infinitesimal transformations $\delta A^a_{\mu} = \epsilon \partial_{\mu} \chi^a$ we obtain
\begin{align}
J^{\mu}_{\chi} = &- F^{\mu \nu}_{a} \partial_{\nu} \chi^a + \xi_{ab} \partial_{\nu} A^{a \nu} \partial^{\mu} \chi^{b} \nonumber\\
&+ j^{a \mu} \chi_{a} + M_{ab} A^{a \mu} \chi^{b}.
\end{align}
This expression can be rewritten as
\begin{align}
    J^{\mu}_{\chi}  = &- \partial_{\nu} \left( F^{\mu \nu}_{a} \chi^a \right) + \xi_{ab} \partial_{\nu} A^{a \nu} \partial^{\mu} \chi^{b}\nonumber \\
    &+ \partial_{\nu} F^{\mu \nu}_a \chi^a + j^{a \mu} \chi_a + M_{ab} A^{a \mu} \chi^b,
\end{align}
which has the form of a divergence of a superpotential (first term) plus additional terms. Once we evaluate this current on-shell we find 
\begin{align}
J^{\mu}_{\chi}  |_{\textrm{on-shell}}= &- \partial_{\nu} \left( F^{a \mu \nu} \chi_a \right)\nonumber \\
 &+\xi_{ab} \left( \partial_{\nu} A^{a \nu} \partial^{\mu} \chi^b - \chi^b \partial^{\mu} \partial_{\nu} A^{a \nu} \right).
 \label{charges}
 \end{align}
As in the electrodynamics case \cite{barcelo2016}, these symmetries are physical symmetries and not gauge. Actually, they are equivalent to the local symmetries discussed in the electrodynamics case for a collection of real fields $\varphi^a = \partial_{\mu} A^{a \mu}$. This can be made explicit by noticing that, under the hypothesis of a conserved matter current, the divergence of the equations of motion
\begin{equation}
\left( \xi_{ab} \Box + M_{ab} \right) \partial_{\mu} A^{a \mu} = 0,
\end{equation}
are always source-free Klein-Gordon equations for the scalar fields $\varphi^a$. Thus, the local symmetries we have introduced  carry nontrivial Noether charges. Indeed, they can be seen to correspond to the Fourier components of the free field expansion \cite{barcelo2016}.

If we restrict ourselves to the subspace $\mathcal{U}$ given by the set of fields obeying
\begin{equation}
\varphi^a = \partial_{\mu} A^{a \mu} = 0,
\label{subspaceU}
\end{equation}
which is quite natural since there are no Lorentz invariant sources that might produce excitations on this scalar sector of the theory, the local transformations  \eqref{linearizedgauge} and \eqref{condition_gauge} become gauge symmetries as long as the matrix $M_{ab}$ identically vanishes, $M_{ab} =0$. 
That is, the Noether charges in (\ref{charges}) all become zero.
This is completely analogous to the electrodynamics case, where we required the mass to vanish for the gauge symmetry to emerge. This is because the gauge transformations (\ref{linearizedgauge}) do not leave the subspace $\mathcal{U}$ invariant
unless the mass matrix $M_{ab}$ is equal zero, since they have the form (see \eqref{linearizedgauge} and \eqref{condition_gauge})
\begin{equation}
\varphi^a \rightarrow \varphi^a - \left( \xi^{-1} \right)^{ab} M_{bc} \chi^c \notin \mathcal{U}.
\end{equation}
It is just for $M_{ab} = 0$ that we have
\begin{equation}
\varphi^a \rightarrow \varphi^a,
\end{equation}
under the transformations (\ref{linearizedgauge}). Moreover, as we have mentioned, once we restrict ourselves to that subspace, the Noether charges associated with these transformations (\ref{charges}) become trivial as it should happen:
\begin{align}
Q_{\chi} &= \int_{\Sigma} d^3x\, \partial_{\mu} \left( F^{a 0 \mu}  \chi_a \right) \nonumber\\
&= \int_{\partial \Sigma_{\infty}} \sqrt{\gamma}\,d ^2x\, n_{\mu}  F^{a 0 \mu}  \chi_a = 0,
\end{align}
where $\Sigma$ is a generic $t=\textrm{constant}$ spacelike surface, $\partial \Sigma_{\infty}$ is the boundary of $\Sigma$ whose normal vector we call $n^{\mu}$, and $\gamma$ is the corresponding induced metric on $\partial \Sigma_{\infty}$. Here we have used Gauss theorem and the last equality follows immediately from choosing appropriate boundary conditions to exclude unphysical solutions with  field strength at infinity.
To summarize, we have a subset $\mathcal U$ of solutions selected by $\varphi^a=0$ that, when $M_{ab}=0$, is invariant under symmetry transformations \eqref{linearizedgauge} and such that the Noether charges identically vanish in it, although not in the whole set of solutions. So, we have precisely all the conditions described in the previous section for the emergence of gauge symmetries. Actually, we can identify these emergent gauge symmetries with the linearization of Yang-Mills gauge symmetries, which constitute a gauge theory whose gauge group is $U(1)^{\otimes N}$. In fact, the analysis up to now is completely equivalent to that in~\cite{barcelo2016}, but now having several copies of vector fields.

We have seen how linearized Yang-Mills theory in the Lorenz gauge has emerged from a theory without this gauge symmetry. The emergence of a linearized gauge symmetry is equivalent to the decoupling of degrees of freedom that would correspond to ghosts, and therefore renders classical instabilities that would be present otherwise irrelevant. This will be even more clear when dealing with the whole nonlinear theory in the following section, where we will apply a bootstrapping procedure. For the moment, let us discuss in this linearized framework two observables that are fundamental in the theory: the energy-momentum tensor and the Yang-Mills current. We will omit the matter content and focus on the contribution from the $A^{a}_{\mu}$ fields exclusively in the rest of this section.

Applying Hilbert's prescription, we can find the symmetric energy-momentum tensor $T_{\mu \nu}$ which agrees with the one obtained following Belinfante's prescription \cite{belinfante1940,rosenfeld1940}
\begin{align}
T_{\mu \nu} = &-F^{a}_{\mu \rho} F^{\rho}_{a \nu}   - \frac{1}{4}\eta_{\mu \nu}F_{a \rho \sigma} F^{a \rho \sigma} + \frac{1}{2} \eta_{\mu \nu}  \xi_{ab} \partial_{\rho} A^{a \rho} \partial_{\sigma} A^{b \sigma}\nonumber \\
&+2 \xi_{ab} A^{a}_{( \mu} \partial_{\nu)} \partial_{\rho} A^{b \rho} - \xi_{ab} \eta_{\mu \nu} \partial_{\rho} \left( A^{a \rho} \partial_{\sigma} A^{b \sigma} \right).
\end{align}
The projection of $T_{\mu \nu}$ onto the subspace $\mathcal{U}$ is clearly built out of the tensor $F^{a}_{ \mu \nu}$, being by construction invariant under the transformations   \eqref{linearizedgauge}. Therefore, the stress-energy tensor is unable to tell the difference between elements of the equivalent classes of solutions within~$\mathcal{U}$.

On the other hand, we can build  Yang-Mills currents in the system. These currents are associated with rotations in the internal space represented by the latin index $a$. Starting from the action $S_0$ given in (\ref{eq:zeroact}), we can see that, for general $\xi_{ab}$ and $M_{ab}$, there are no such symmetries in the complete theory. However, under certain restrictions we notice that the action is invariant under the rigid transformations
\begin{equation}
A^{a \mu} \rightarrow A^{a \mu} + f^{abc} A^{\mu}_b \zeta_c,
\label{rigidtransf}
\end{equation} 
where $\zeta^{c}$ is an arbitrary set of constants and the constants $f^{abc}$ need to verify 
\begin{equation}
f^{abc} = f^{[ab]c}.
\label{antisymmetry}
\end{equation}
Let us stress that this condition is weaker than the Jacobi identity for $f^{abc}$. The necessary restrictions are the fields being massless, $M_{ab}=0$,  and $\xi_{ab}$ satisfying the following condition (written for simplicity in terms of $\Xi^{acd} = \xi^{a}_{b} f^{bcd}$):
\begin{equation}\label{eq:conscond}
    \Xi^{acd} = \Xi^{[ac]d}.
\end{equation}
This condition guarantees that the term in the action proportional to $\xi_{ab}$ is invariant under the transformations   \eqref{rigidtransf}. This condition is satisfied for instance if
\begin{equation}\label{eq:conscondres}
    \xi_{ab}=\lambda \delta_{ab}.
\end{equation}
There may be more general situations in which Eq. \eqref{eq:conscond} is satisfied. However, for simplicity we will restrict our analysis below to theories in which $\xi_{ab}$ satisfies Eq. \eqref{eq:conscondres}.

According to Noether's theorem, associated with any symmetry of this sort we have a conserved current, which reads
\begin{equation}
J^{(1) a \mu} = f^{bca} \left[ F^{\mu \nu}_b A_{c \nu}
-\lambda  (\partial_{\nu} A^{b\nu}) A^{c\mu} \right].
\label{current_firstorder}
\end{equation}
It is instructive to check that this current is conserved upon imposing the equations of motion. Taking the divergence of Eq. \eqref{current_firstorder} and suitably grouping the terms we find
\begin{align}
\partial_{\mu} J^{(1) a \mu} &= f^{bca} \left( \partial_{\mu} F^{\mu \nu}_b - \lambda \partial^{\nu} \partial_{\mu} A^{b \mu} \right) A_{\nu c}\nonumber \\
&+ f^{bca}\left(  F^{\mu \nu}_{b} \partial_{\mu} A_{c \nu} - \lambda \partial_{\nu} A^{\nu}_{b} \partial_{\mu} A^{\mu}_{c} \right).
\end{align}
We have four terms in this divergence. The first two terms vanish on-shell, i.e., imposing the equations of motion, Eq. \eqref{linearYM}, with $M_{ab} = 0$ and $j^{a \mu} = 0$. The third term vanishes due to the symmetry structure of the Lorentz and internal indices. Since the tensor $F^{\mu \nu}_{b}$ is antisymmetric in its Lorentz indices $F^{\mu \nu}_{b} = F^{[\mu \nu]}_b$, its contraction with another tensor just picks the antisymmetric part of such tensor. In our case, the antisymmetric part of $\partial_{\mu} A_{\nu c}$ is proportional to $F_{\mu \nu c}$. This means that the third term can be written as
\begin{equation}
    \frac{1}{2} f^{bca} F^{\mu \nu}_{b} F_{\mu \nu c} = \frac{1}{2} f^{[bc]a} F^{\mu \nu}_{(b|} F_{\mu \nu |c)} =0,
\end{equation}
where we have taken into account the antisymmetry of $f^{bca}$ in its two first indices, and the fact that the contraction of the Lorentz indices of the $F$-tensors is symmetric in the internal indices. Finally, the fourth term is manifestly the same kind of contraction of a symmetric object with an antisymmetric one
\begin{equation}
    f^{bca} \partial_{\nu} A^{\nu}_b \partial_{\mu} A^{\mu}_{c} = f^{[bc]a} \partial_{\nu} A^{\nu}_{(b|} \partial_{\mu} A^{\mu}_{|c)} =0.
\end{equation}
Thus, we have proved that the current given by Eq. \eqref{current_firstorder} is conserved on-shell.

Here we find the first difference with the respect to a single vector field. Having several copies of vector fields allows us to prescribe interactions between them, which in turn allows for the $A^{a}_{\mu}$ fields to become charged themselves. As we will see, there are many different possibilities to prescribe interactions, essentially as many as different Lie algebras of dimension $N$. But the important observation at this stage is the following: even restricting to the constraint surface $\varphi^a = \partial_\mu A^{a \mu}=0$, which eliminates the second term in (\ref{current_firstorder}), none of the currents that we can build are invariant under the emergent gauge transformations. Notice however that the charges $Q^a$ obtained by integrating the zero component of the currents are indeed invariant under these transformations and nontrivial.   

Therefore, these charges are in principle observables that one could use to distinguish between different solutions within the equivalence classes associated with emergent gauge symmetries. The presence of these charges can be interpreted as traces of the complete theory, recalling that it is not a gauge theory, \textit{ab initio}. We will continue the discussion of this important issue in the discussion section. To finish the section, let us just mention that the currents (\ref{current_firstorder}) are the ones that we will use to couple the fields $A^{a}_{\mu}$ with themselves at first order, on the way toward building a proper Yang-Mills theory.

\section{Bootstrapping Lorentz-invariant vector fields}
\label{section4}

Up to now we have seen that if in a complicated theory there is a low-energy regime with an emergent Lorentz symmetry for a collection of Lorentz-invariant vector fields, then one will immediately deduce that the system develops for free, under the massless assumption, the appearance of emergent gauge symmetries. But once the linearization of Yang-Mills theory has emerged in our system, it is natural to analyze whether this emergence can be extended (perhaps in a unique way) to the nonlinear regime. This question will be explored in this section.

\subsection{General idea and bootstrap procedure}

To answer this question, let us assume for simplicity that we are dealing with the theory in vacuum (vanishing source currents, $j^{a \mu} =0$). Including the matter content back in the equations of motion will be straightforward and we will comment on that in Sec. \ref{sec:matter}.

Let us recall that the projection that we have made onto the subspace $\mathcal{U}$ for which the gauge invariance emerges is defined by the constraints (\ref{subspaceU}), i.e., $\partial_{\mu} A^{a \mu} =~0 $. These constraints emerged when analyzing the equations of motion of the linear theory, and can be therefore considered on-shell from this perspective. Given that the bootstrapping procedure aims at deriving the action of a suitable nonlinear completion of a given linear theory, the most straightforward procedure is not to include these constraints as a part of the bootstrapping procedure (otherwise, the problem would be equivalent to a gauge-fixed version of the one considered by Deser in \cite{deser1970}). In other words, we will be analyzing the bootstrapping of Lorentz-invariant linear theories of vector fields with no gauge invariance \emph{a priori}. 

In practice, the starting point for the bootstrapping procedure is the action $S_0$ defined in Eq.~\eqref{eq:zeroact} with $\xi_{ab}~=~\lambda\delta_{ab}$, $M_{ab}=0$, and $j^{a \mu} =0$, namely
\begin{equation}\label{eq:zeroboots}
S_0 = \int d^4x  \left[ -\frac{1}{4}F^{a}_{ \mu \nu} F_{a}^{\mu \nu} +\frac{\lambda}{2}   (\partial_{\mu} A_{a}^{ \mu} ) (\partial_{\nu} A^{a \nu}) \right],
\end{equation}
Let us recall that $F^a_{\mu\nu}$ was defined in Eq. \eqref{eq:lsdef}. The additional condition that we will have in this perturbative reconstruction of the theory will be preserving the order of the equations of motion: We want the most general theory compatible with having second order differential equations.

The case $\lambda=0$ corresponds to the usual description of the bootstrapping of Yang-Mills theory. In that case, for consistency reasons, any kind of self-interactions that we add to the equations of motion of the free theory need to be introduced via a conserved current \cite{deser1970,ortin2010}. The idea is to consider a small coupling constant $g$ and introduce a conserved current present on the free theory on the right hand side, that is any of the currents $J^{(1) a \mu}$ in (\ref{current_firstorder}) without the second term (as the latter vanishes when $\lambda=0$): $J^{(1) a \mu} = f^{bca}F^{\mu \nu}_{b} A_{c \nu}$. Any current on this set is bilinear in the $A^{a \mu}$ fields, and will introduce the first nonlinearities in the theory. In other words, we will consider that $\mathcal{J}^{a \mu} = g J^{(1) a \mu} $ to first order. While for $\lambda\neq0$ one does not need to consider conserved currents, our discussions of the electrodynamics case in \cite{barcelo2016} and linearized Yang-Mills in the previous section strongly suggest that this will be a necessary ingredient for the emergence of gauge symmetries at the nonlinear level. In any case, it has been discussed in \cite{padmanabhan2004,ortin2010,barcelo2014} that, even when considering conserved currents, there is no unique current that can be chosen at this stage of the procedure in a natural way. A complementary approach toward making a consistent nonlinear extension of the linear spin-1 and spin-2 theories was put forward by Ogievetsky and Polubarinov \cite{Ogievetsky1962,ogievetsky1965b,ogievetsky1965} (see also the brief review in \cite{Ivanov2016}).

This ambiguity in the choice of current comes from the possibility of adding boundary terms to the action which translate into identically conserved additional pieces for the current computed via Noether's procedure. For the first nontrivial interaction $J^{(1) a \mu}$, such pieces come from the possible boundary terms that we can add to the quadratic action $S_0$ while keeping its linear, second order character and containing at most quadratic terms in the time derivatives,
\begin{equation}\label{eq:bterms}
    S_{0,B} = \int d^4x  Q^{ab}_{\ \ \mu \nu \rho\sigma} \partial^{\mu} \left( A^{a \nu} \partial^{\rho} A^{b \sigma} \right),
\end{equation}
where $Q^{ab}_{\ \ \mu \nu \sigma \rho}$ is constructed with the tensorial quantities available, namely $\delta^{ab}$ and $\eta_{\mu\nu}$:
\begin{equation}\label{eq:qansatz}
    Q^{ab}_{\ \ \mu \nu \rho \sigma}=\delta^{ab} B \left(\eta_{\mu\sigma}\eta_{\nu\rho} - \eta_{\mu\nu}\eta_{\rho\sigma}\right),
\end{equation}
with $B$ an arbitrary constants. The contribution of this boundary term to $J^{(1) a \mu}$ can be straightforwardly computed, adding to the current in Eq. (\ref{current_firstorder}) terms of the form
\begin{align}
J^{(1,B)a}_{\mu} &=\left(f^{dca}Q^{db}_{\ \ \mu\nu\rho\sigma}+f^{dca}Q^{bd}_{\ \ \rho\sigma\mu\nu} \right.\nonumber\\
&+ \left.f^{dba}Q^{cd}_{\ \ \mu\nu\rho\sigma}-f^{dca}Q^{bd}_{\ \ \mu\sigma\rho\nu}\right)A^{c\nu}\partial^\rho A^{b\sigma}.
\label{boundary_term}
\end{align}
When using Eq. \eqref{eq:qansatz} this expression is simplified to
\begin{equation}
J^{(1,B)a}_{\mu} =Bf^{bca}\left(\eta_{\mu\sigma}\eta_{\nu\rho}-\eta_{\mu\nu}\eta_{\rho\sigma}\right)A^{c\nu}\partial^\rho A^{b\sigma}.
\end{equation}
It is straightforward to check that the tensor $f^{bcd}\left(\eta_{\mu\sigma}\eta_{\nu\rho}-\eta_{\mu\nu}\eta_{\rho\sigma}\right)$ in the equation above is antisymmetric under the exchanges $\mu\leftrightarrow\rho$, $\nu\leftrightarrow\sigma$ and $b\leftrightarrow c$ independently, which in particular implies that $J^{(1,B)a}_{\mu}$ is identically conserved. These ambiguity is inherent to the bootstrapping procedure, as this procedure by itself does not prefer one choice of current or another. The specific current that we use (namely, the specific value of $B$) needs to be given as an input.

In summary,  we will consider the conserved source given in Eq. \eqref{current_firstorder} (with possible contributions from boundary terms added to it) as the source of the equations of motion at first order, even if such conservation is not required from the perspective of bootstrapping. Hence, this represents an additional assumption of our construction, that at this stage can be motivated by the invariance under the rigid transformations \eqref{rigidtransf}, which singles out this conserved current. We will provide additional motivation for this choice below, once the implications that it has for the bootstrapping procedure become clear. We will also keep in mind the inherent ambiguity in the choice of a current that is associated with boundary terms, as discussed above, and eventually explain how to deal with it.

Now, to be able to derive these currents from an action principle, we need to add a term of order $g$ to the action, $S = S_0 + g S_1$, such that
\begin{equation}\label{eq:boots1st}
J^{(1) a \mu} =  \frac{\delta S_1}{\delta A_{a \mu}}.
\end{equation}
Adding this new term to the action will modify the current obtained via Noether's procedure by a term of order $\order{g^2}$, in addition to possibly imposing some consistency conditions on the original symmetry that we identified in the free theory. Thus, we will have $\mathcal{J}^{a \mu} =g  J^{(1) a \mu} + g^2 J^{(2) a \mu}$. We have again the presence of ambiguities in the choice of $J^{(2) a \mu}$. These come now from the possibility of adding additional boundary terms to $S_1$ of the same order and containing the same number of fields it contains. This new piece added to the action will require adding a term of order $\order{g^2}$ to the action $S = S_0 + g S_1 + g^2 S_2$ such that
\begin{equation}
J^{(2) a \mu} =  \frac{\delta S_2}{\delta A_{a \mu}}.
\end{equation}
This will iteratively generate an action of the form
\begin{equation}
S = \sum_{n = 0}^{\infty} g^n S_n,
\end{equation}
where at each step we may produce additional constraints and the functionals $S_n$ are built in order to match the contribution to the current generated by the $n-1$ term. At each step, we produce ambiguities with the same nature as the ones we have already discussed, as at each order we can add additional boundary terms to the action that translate into additional pieces for the current.

At the end of the day, we will have that we can break the action into a free and an interacting part $S = S_0 + S_I$, such that the variation of the interacting part will give us the whole current to which we couple the free term 
\begin{equation}
\frac{\delta S_I}{\delta A_{a \mu}} = \mathcal{J}^{a \mu}.
\label{full_current}
\end{equation}
Although when applied to gravity this procedure requires the sum of an infinite series \cite{barcelo2014}, we will see that for the Yang-Mills theory this procedure stops at order $\order{g^2}$ and $S_n \equiv 0$, for $n>2$. As an additional part of the bootstrapping procedure we will also discuss how this affects the charged matter sector. Then in the next section we will discuss what happens when applying the bootstrapping to the unconstrained theory.

\subsection{Explicit integration and summation of the series}

Let us begin the process with the action $S_0$ given in Eqs. \eqref{eq:zeroboots} and \eqref{eq:bterms}. We have already discussed that we will select the current in Eq. \eqref{current_firstorder}, plus possible contributions from boundary terms \eqref{boundary_term}, as the source $J^{(1)a\mu}$ at first order. Before discussing the role of boundary terms, let us focus on the term in this current that is proportional to $\lambda$, namely
\begin{equation}
    J^{(1)a\mu}_{(\lambda)}= \lambda f^{bca} \partial_{\nu} A_{b}^{ \nu} A_{c}^{ \mu}.
\label{g-current-1}
\end{equation}
As the equation to be solved in order to obtain $S_1$, namely Eq. \eqref{eq:boots1st}, is a linear equation, we can consider independently the piece of the action $S_{1(\lambda)}$ that leads to the piece of the current above under its variation. As it is discussed in Appendix \ref{sec:appendix}, there is no choice of $S_{1(\lambda)}$ that can lead to this current. This implies that the bootstrapping procedure can be completed for the current associated with the rigid symmetries \eqref{rigidtransf} if and only if $J^{(1)a\mu}_{(\lambda)}$ vanishes, which generically leads to the same condition that we identified at the linear level when discussing the emergence of gauge symmetries, namely $\partial_\mu A^{a\mu}=0$. 

This condition now appears as a requirement that must be satisfied in order to be able to find an action (and therefore, to proceed with the bootstrapping) for the choice of current at first order. This constraint appears then as a structural requirement of the bootstrapping procedure. As in the analysis of the linear theory described in Sec. \ref{section_2}, this constraint ensures that the scalar degrees of freedom encoded in $A^{a\mu}$ decouple. The resulting theory is by construction equivalent to Yang-Mills in the Lorenz gauge and is therefore free of ghosts and other pathologies expected in the absence of gauge symmetries. Let us stress that the ambiguity associated with boundary terms cannot change this conclusion, as shown in Appendix \ref{sec:appendix}. In physical terms, this implies that this decoupling is a robust condition that must be satisfied for every conserved current associated with the symmetry under the transformations \eqref{rigidtransf}.

Hence, in the following we will assume that the fields are divergenceless, which we will implement in the action through a Lagrange multiplier to be added to $S_1$, which we can write without loss of generality (more details are provided in Appendix \ref{sec:appendix}) as
\begin{equation}
S_1=\int d^4x \, P^{bca}_{\ \ \ \ \mu\nu\rho\sigma} \partial^\mu A^{b \nu} A^{c \rho} A^{a\sigma},
\label{P_action}
\end{equation}
where
\begin{align}\label{eq:pdef}
     P^{abc}_{\ \ \ \ \mu \nu \rho \sigma}\ = & \eta_{\mu \nu}\eta_{\rho \sigma}  \alpha \left(f^{abc}-f^{cab} \right)  \nonumber\\
    +&\eta_{\mu\rho}\eta_{\nu\sigma}\left(\beta_1f^{abc}+\beta_2f^{cab}+\beta_3f^{bca}\right) \nonumber\\
    -&\eta_{\mu\sigma}\eta_{\nu \rho}\left(\beta_2f^{abc}+\beta_1f^{cab}+\beta_3f^{bca}\right).
\end{align}
The condition that this action leads to $J^{(1)a\mu}$ when variations with respect to $A^{a\mu}$ are considered implies the following algebraic relation:
\begin{align}
        P^{bca}_{\ \ \ \ \sigma\nu\rho\mu}-P^{abc}_{\ \ \ \ \sigma\mu\nu\rho} &= f^{bca}(\eta_{\mu\sigma}\eta_{\nu\rho}-\eta_{\mu\nu}\eta_{\rho\sigma})\nonumber \\
        &+Bf^{bca}(\eta_{\mu\nu}\eta_{\rho\sigma}-\eta_{\mu\rho}\eta_{\nu\sigma}).
        \label{algebraic_system}
\end{align}
This translates into an incompatible set of equations for the parameters $(\alpha,\{\beta_i\}_{i=1}^3,B)$. Thus, the system of equations that follows from \eqref{algebraic_system} has no solution as long as we do not impose further constraints on the components of the tensor $f^{abc}$. This system of equations has no solution as long as we do not impose further constraints on the components of the tensor $f^{abc}$ that reduce its number of independent components, which may result in a compatible system. A natural condition to impose is full antisymmetry of $f^{abc}$. The naturalness of this choice stems from the fact that, once we consider the action \eqref{P_action}, requiring the constants $f^{abc}$ to obey the Jacobi identity
\begin{equation}
f^{ade}f^{bc}_{~~d} + f^{bde}f^{ca}_{~~d} + f^{cde}f^{ab}_{~~d} = 0
\label{jacobi}
\end{equation}
ensures that the transformations \eqref{rigidtransf} remain symmetries of the theory with action $S_0 + g S_1$. Hence, Eq.~\eqref{jacobi} is an additional consistency condition to the antisymmetry in Eq.~\eqref{antisymmetry}, being these imposed to guarantee that the rigid transformation~\eqref{rigidtransf} is a symmetry at first and zeroth order, respectively. Without this additional consistency condition, the iterative procedure cannot be continued. Taking into account that the tensor $f^{abc}$ obeys the Jacobi identity and we have an Euclidean metric $\delta^{ab}$ in that space, we conclude that $f^{abc}$ needs to be completely antisymmetric. Hence, $f^{abc}$ can be understood as the structure constants of a compact semisimple Lie algebra~\cite{georgi1999}. Our latin indices run from $1$ to $N$, meaning that the dimension of the Lie algebra and its corresponding group is the number $N$ of independent generators. If we want the structure constants to close an $\mathfrak{s} \mathfrak{u}(M)$ algebra, for instance, we would additionally require $N= M^2-1$. Imposing this ansatz with the full antisymmetry of $f^{abc}$, the tensor $P^{bca}_{\ \ \ \ \sigma\nu\rho\mu} $ reduces to 
\begin{equation}
        P^{bca}_{\ \ \ \ \sigma\nu\rho\mu} = 
    \beta f^{abc} \left( \eta_{\mu\sigma}\eta_{\nu\rho}  - \beta \eta_{\mu\nu}\eta_{\rho\sigma} \right).
    \label{antisymmetric_ansatz}
\end{equation}
Plugging this ansatz in the algebraic relation \eqref{algebraic_system}, we obtain the a compatible system of equations for the parameters $(\beta, B)$ whose unique solution is $\beta = B = -1$.

We have thus obtained
\begin{equation}
S_1 = - \frac{1}{2} \int d^4x \left[ f^{abc} F_{a \mu \nu} A^{\mu}_b A^{\nu}_c  - \vartheta_a \partial_{\mu} A^{a \mu} \right], 
\label{S1}
\end{equation}
where we have added explicitly the Lagrange multipliers $\vartheta^a$ that enforce the required constraints on the fields $A^{a\mu}$ for the self-consistency of the bootstrapping. As already noticed by Deser \cite{deser1970}, the direct current $J^{(1) a \mu}$ in Eq. (\ref{current_firstorder}) does not lead to the action in Eq. \eqref{S1}. The ambiguity in the definition of the current due to boundary terms in $S_0$ must be taken into account in order to provide additional contributions to the current necessary for the bootstrapping procedure to work. We notice that these general ambiguities were not emphasized enough in the past, as it was considered that the bootstrapping procedure did not require to take these ambiguities into account. The reason for this was that the fist-order formalism used by Deser in \cite{deser1970} leads to Yang-Mills for a trivial choice of these boundary terms, while this is no longer true for the second-order formalism defined using the vector fields $A^{a\mu}$. However, this does not imply that the second-order formalism cannot be used for the bootstrapping procedure, as we have seen explicitly that the self-consistency of the iterative procedure is enough to select the necessary boundary terms so that there is a unique solution, up to a choice of a semisimple Lie-Algebra of the same dimension as the number of fields involved in the construction. The important role played by boundary terms in the gravitational case was discussed in \cite{padmanabhan2004,Butcher2009,barcelo2014}. It would be interesting to have a clear understanding of the similarities and differences between Yang-Mills and gravitational theories from this perspective.

In the next step, the first term from $S_1$ produces a contribution to the current given by
\begin{equation}
J^{(2) \mu}_{a} = f^{bcd} f_{bea} A^{\mu}_{d} A^{\sigma}_{c} A_{\sigma}^{e}.
\end{equation}
The second term from $S_1$ gives a current of the form 
\begin{equation}
J^{ a \mu}  =  \frac{1}{2} f^{a b c} \vartheta_b A^{\mu}_c,
\label{lagrange_multiplier_current}
\end{equation}
which cannot be derived from variational principle for $A^{a \mu}$. The only term that might give rise to a current proportional to $A^{a \mu}$ is a term of the form $A^{a}_{\mu} A^{b \mu}$. However, we would need to contract this with $f^{a b c}$, which is antisymmetric and hence this term would identically vanish. Thus, this second contribution to the current would apparently break the bootstrapping procedure. To avoid breaking the bootstrapping procedure, we need to add a boundary term that cancels the contribution from~\eqref{lagrange_multiplier_current} once it is evaluated on-shell. The equations of motion of the constrained theory (after imposing the transverse condition) are such that the Lagrange multipliers need to be constants.

Notice that the equations of motion when we add the Lagrange multipliers are the following 
\begin{align}
    \partial_{\mu} F^{a \mu \nu} - \partial^{\nu} \partial_{\mu} A^{a \mu} - g \partial^{\mu} \theta^a = J_1^{a \mu} + \order{g^2}, \nonumber \\
    \partial_{\mu} A^{a \mu} = 0,
\end{align}
where we have explicitly indicated that the first equation is correct up to order $g^2$ (since it is obtained through a bootstrapping procedure). Taking the divergence of the first equation, using the second equation and additionally taking into account that the current $J_1^{a \mu}$ is conserved up to order $g^2$, we have that to the desired order the equations for the Lagrange multipliers $\theta^a$ reduce to a set of sourceless wave equations
\begin{equation}
    \partial_{\mu} \partial^{\mu} \theta^a = 0.
\end{equation}
The same arguments that we have put forward below Eq.~\eqref{subspaceU} apply to the $\theta^a$ fields here. Hence, the Lagrange multiplier reduce to be constants on-shell. If we ask for them to vanish asymptotically, as we do for the $A^{a \mu}$ fields, these zero-modes need to be equal to zero. However, for our purposes it is irrelevant what the actual value of the zero-modes is.

Hence, on-shell we have that $\vartheta^a = \kappa^a \in \mathbb{R}$. Thus, the addition of a boundary term of the form 
\begin{equation}
    S_{b} = - \frac{1}{2}  g \int d^4x \kappa_a \partial_{\mu} A^{\mu a},
    \label{boundary_term}
\end{equation}
identically cancels the contribution from Eq.~\eqref{lagrange_multiplier_current}. Notice that we have not added any other boundary terms to $S_1$ to build $J^{(2) \mu}_a$, except the one from Eq.~\eqref{boundary_term}. This choice is precisely the choice that fulfills our criteria of providing a nonlinear theory obtained through a bootstrapping procedure that implements a deformation of the original gauge symmetry and preserves the number of degrees of freedom. 

For the sake of completeness, let us mention the existence of an alternative procedure that, while being equivalent on-shell, is arguably simpler. In this alternative procedure, the constraint $\partial_\mu A^{a\mu}$ is implemented quadratically instead of linearly, using a scalar Lagrange multiplier that enforces $\partial_\mu A^{a\mu}\partial_\nu A^{\nu}_a=0$. In this implementation, the corresponding contributions to the current vanish on-shell without the necessity of further considerations.

The action that leads to this current is
\begin{equation}
S_2 = \frac{1}{4} \int d^4x f^{bcd} f_{bea} A^{\mu}_{d} A^{\sigma}_{c} A_{\sigma}^{e} A^{a}_{\mu}.
\end{equation}
The iterative procedure happens to stop here. The reason is that the $S_2$ term does not contribute to the current $\mathcal{J}^{a \mu}$: It contains no derivatives of the $A^{a}_{\mu}$ fields and it is strictly invariant under the rigid transformations \eqref{rigidtransf}, thus making its contribution to the current computed via Noether's procedure identically zero.

The final action $S=S_0+gS_1+g^2S_2$ can then be written, after conveniently reorganizing the terms, as
\begin{align}
S = \int d^4x  \bigg [ &-\frac{1}{4}\mathcal{F}^{a}_{ \mu \nu} \mathcal{F}^{ \mu \nu}_{a} + \vartheta_a \partial_{\mu} A^{\mu a} \\
& +\frac{\lambda}{2}  (\partial_{\mu} A_{a}^{ \mu} ) (\partial_{\nu} A^{a \nu})   \bigg ],
\label{nl_action}
\end{align}
where $\vartheta_a$ are Lagrange multipliers and the nonlinear field strength tensor $\mathcal{F}^{a}_{ \mu \nu}$ has the Yang-Mills field form 
\begin{equation}
\mathcal{F}^{a}_{\mu \nu} = 2 \partial_{[\mu} A^a_{\nu]} + g  f^{bca}A_{b\mu} A_{c\nu}.
\end{equation}
In this way we have found the free Lagrangian of a non-Abelian Yang-Mills theory in the Lorenz gauge. The conditions (\ref{antisymmetry}) and (\ref{jacobi}) on the coefficients $f^{abc}$ that determine the symmetry transformations prescribe the structure of a particular Lie algebra (they are the structure constant of the corresponding algebra). In the process of making the theory nonlinear we are selecting a particular form of Lie algebra among those of dimension $N$. Then, we can write $A^{\mu} = A^{a \mu} T_{a}$, with $T^a$ representing the generators of a semi-simple and compact Lie algebra which satisfy the algebraic relations $[T^a,T^b] = i f^{abc} T_c$.

The equations of motion resulting from this action are invariant under the deformation of the linear transformations \eqref{linearizedgauge}. These transformations are given by the exponentiation of the transformations 
\begin{equation}
A^{a}_{ \mu} \rightarrow A^{a}_{ \mu} + \partial_{\mu} \chi^a + g f^{abc} A_{b \mu} \chi_c,
\label{gaugenl}
\end{equation}
with $\chi^a$ obeying the following constraints that guarantee that the transformations above do not make the fields leave the subspace $\mathcal{U}$ defined by the Lorenz condition:
\begin{equation}
\Box \chi^a + g f^{abc}   A^{\mu}_{b} \partial_{\mu} \chi_c =0.
\end{equation}
For the purposes of doing perturbative calculations, the propagator associated with the quadratic part of the action can be worked out directly as that of $N$ free Maxwell fields within the subspace of transverse fields; See~\cite{bhattacharjee2013} for an explicit and straightforward derivation of such propagator.

\subsection{Inclusion of matter}\label{sec:matter}

The coupling to the matter content can also be obtained perturbatively via another bootstrapping process, as we have advanced in the previous section. Let us make explicit how this is done, for instance, for a set of fermionic fields as if we aimed at constructing an emergent QCD theory. Let us represent the label corresponding to diferent flavors of fermions $\psi^i$ with latin indices $i,j,k$ and assume that those indices run from $1$ to $M$.
In the logic of our iterative construction we first use as source of the $A^{a\mu}$ fields a free fermionic current which has the form
\begin{equation}
j^{a \mu} = \bar{\psi}^{i} \gamma^{\mu} \tilde{T}^a_{ij} \psi^{j},
\end{equation} 
where the matrices $\gamma^{\mu}$ are Dirac gamma matrices and $\tilde{T}^{a}_{ij}$ are certain unspecified matrices. The free equation for the fermionic fields guarantees that this current is conserved no matter   what $\tilde{T}^{a}$ we use. 
To see this, we just need to take the divergence of this current and use the equations of motion at zero order. 
Under these conditions, the decoupling of longitudinal degrees of freedom is ensured and we have the emergent gauge symmetries that we already had in the electrodynamics example.
 
We want to obtain the equations for $A^{a\mu}$ from a Lagrangian   in such a way that the source of the field includes both  the first order nonlinearities of the fields $A^{a\mu}$ and this fermionic source current. In a first iteration, we therefore prescribe an action of the form $S_0+S_1+S_{f1}$ (see (\ref{S1})) with 
\begin{equation}
S_{f1} = \int d^4 x  \left[ \bar{\psi}^j \gamma^{\mu} \left(i \delta_{ij} \partial_{\mu} -  q_a \tilde{T}^a_{ij} A_{\mu}^{a} \right) \psi^{i} \right].
\end{equation}
Here the constants $q_a$ represent the charges associated with each of the $U(1)$ copies of the system that we found in the linearized theory. From a physical point of view, it implies that the fermionic source not only affects the fields $A^{a\mu}$ but that in return they also affect the fermionic fields. Then, the resulting new equation for the fermions no longer need to fulfill an exact conservation condition, only conservation up to $\order{q_a}$. This is reasonable, as potentially only a sum of fermionic plus Yang-Mills currents should be conserved, and moreover, a proper conservation will only appear when closing the bootstrapping procedure. As an aside, let us mention that by choosing the matrices $\tilde{T}^a$ to be commuting matrices, or equivalently, multiples of the identity, one would be able to maintain the conservation of the fermionic current at this first order, but at the cost of killing any possibility for the action to be invariant under rigid rotations of Yang-Mills type. This structure for the fermionic current would be consistent with a theory without the $S_1$ term, that is a linear theory in which a set of noninteracting Maxwell field are coupled to a set of noninteracting fermionic fields. But this is not what we are seeking for here.

The previous action for the fermions has another problem equivalent to that appearing with the action $S_1$ in equation (\ref{S1}) for the field $A^{a\mu}$: As it stands it is not directly invariant under rigid rotations (\ref{rigidtransf}). On the one hand, it is clear that this action can only have the chance to be invariant under rigid rotations if the fermionic fields simultaneously transform as  
\begin{equation}
\psi^i \rightarrow \psi^i - i \chi_a \tilde{T}^a_{ij} \psi^j.
\end{equation}
On the other hand, not all sets of couplings $q_a$ and matrices $\tilde{T}^a_{ij}$ allow for a rigid symmetry, which one needs to continue the bootstrapping procedure. In the case of the $S_1$ action, requiring the existence of a rigid symmetry restricted the form of the coefficients $f^{[ab]c}$ to those closing a Lie algebra (Jacobi property). In the bootstrapping process one has to actively select a specific Lie algebra. It is now clear that the bootstrapping procedure does not select or point toward a concrete one.  Now, requiring the presence of a rigid symmetry in the fermionic sector implies setting all the $q_a$ to a single $g$ and requiring the matrices $\tilde{T}^a_{ij}$ to be precisely fermionic-space representation of the same Lie algebra selected for the $A^{a\mu}$ sector: 
\begin{equation}
[\tilde{T}^a,\tilde{T}^b] = i f^{abc}\tilde{T}^c.
\end{equation}
We will denote these specific matrices $T^a_{ij}$ (without a tilde). Again, unless one forces the theory to follow this specific rule one does not obtain a consistent theory. 

Once we make this choice, there do not appear any more restrictions in the fermionic sector at the next order $\order{g^2}$ and, as such, the bootstrapping process is identical in its next step to the one described in previous sections. The result is that we need to add the matter term
\begin{equation}
S_f = \int d^4 x \left[ \bar{\psi}^j \gamma^{\mu} \left(i \delta_{ij} \partial_{\mu} -  g T^a_{ij} A_{a \mu} \right) \psi^{i} \right],
\end{equation}
to the action (\ref{nl_action}).

Once the iterative process has finished, the final fermionic current plus the Yang-Mills current is the one that is conserved; neither of them is divergenceless separately. Equivalently, we can rephrase this assertion saying that the fermionic current is not conserved but covariantly conserved. For instance, the covariant conservation of the matter current is the necessary condition in the nonlinear theory for the decoupling of degrees of freedom. Moreover, this action is invariant with respect to the infinitesimal gauge transformations (\ref{gaugenl}), if we additionally perform an infinitesimal local transformation of the form 
\begin{equation}
\psi^i \rightarrow \psi^i - i \chi_a(x) \tilde{T}^a_{ij} \psi^j.
\end{equation}
in the fermionic sector.

Thus, we arrive to the conclusion that the result of the bootstrapping process is that in order to have a consistent theory we need all the coupling constants $q_a$ to be the same $g$. Furthermore, we need the matrices $T^a_{ij}$ (that determine the interactions between fermions and $A^{a}_{\mu}$ fields) to be representations of the Lie algebra defined by the selected constants $f^{abc}$. Similar comments apply straightforwardly to other kind of matter coupled to the fields $A^{a}_{\mu}$.

Spin-zero and spin-one matter fields were considered in detail for field theories displaying gauge symmetries \textit{ab initio} in~\cite{basu2018}. We notice that these results can be straightforwardly extrapolated to our framework of emergent gauge theories, as the bootstrapping in the matter sector does not interfere with the bootstrapping in the gauge sector.

\subsection{Gribov copies in this framework}

The standard approach to construct a gauge theory assumes that there are redundancies in our description from the start. Configurations related by gauge transformations, i.e. those with vanishing Noether charges, represent the same physical state.
Gauge symmetries define classes of equivalence within the configuration space of fields. In such situations, the gauge fixing conditions, like the Lorenz gauge
\begin{equation}
\partial_{\mu}A^{a \mu} = 0.
\label{lorenzgauge}
\end{equation}
are introduced in order to choose a representative of each class of equivalence. However, such conditions do their job well if there is a unique representative of those classes that respects the condition. The problem is that for nonlinear gauge theories, the Lorenz gauge does not cross each gauge orbit~\footnote{By gauge orbit of a certain configuration we mean the subspace within field configurations that can be obtained via a gauge transformation, i.e., the subspace of gauge-equivalent configurations to a given one.} once, as was shown by Gribov \cite{gribov1978,gribov1978b}. In general there exist more than one configuration of the fields $A^{a}_{\mu}$ related by gauge transformations, all of them obeying the condition (\ref{lorenzgauge}). They are typically refered to as Gribov copies. 

This is, for instance, problematic from the point of view of defining the quantum theory via path integral techniques. The gauge fixing conditions are needed in order to make sense of the theory by summing over physically inequivalent configuration and the Lorenz gauge condition is typically used because it is explicitly Lorentz invariant. At the perturbative level, this ambiguities are not relevant because we are exploring small deviations in field space from the background solution (which is typically the $A^{a}_{\mu}=0$ configuration, although it could be any other stationary point of the action where we can base our Gaussian perturbative expansion of the theory \cite{coleman1985}). The Lorenz gauge condition is good enough to ensure that there are no Gribov copies around these saddle points but, when exploring the nonperturbative regime of the theory, nothing forbids them to appear. Thus, one enters in conflict with defining the theory nonperturbatively. Gribov actually argued that these ambiguities could have a huge impact in the structure of the quantum theory. He even provided arguments supporting how an appropriate treatment of these features (for example, reducing the path integral to a region absent of Gribov copies, often called a fundamental modular region) could be related to the color confinement characteristic of gauge theories. This is because the restriction of the integration to that region has the effect of generating a linear increase of the interactions between color charges in the deep infrared. Thus, it is a possible mechanism for the explanation of confinement, although a conclusive analysis is not yet available: The theory becomes strongly coupled in that regime and the typical perturbative computations are not reliable \cite{gribov1978,gribov1978b}.

The procedure we have followed to construct a gauge theory is quite different from the standard approach.
In our formalism, we began with a theory that had no physical symmetries and the gauge symmetries emerged after a suitable projection onto a natural subspace of the theory described, precisely, by the Lorenz condition (\ref{lorenzgauge}). The emergent gauge symmetries we refer to are the transformations given by the exponentiation of \eqref{gaugenl}.
This means that these symmetries are the set of transformations acting on the $A_{\mu} = A_{\mu}^a T_a$ fields as
\begin{equation}
A_{\mu} \rightarrow \Omega(x) A_{\mu} \Omega^{-1}(x) + i \Omega(x) \partial_{\mu} \Omega^{-1}(x),
\label{finite_gauge_transf}
\end{equation}
with $\Omega(x) = \exp \left[ i \chi^a(x) T_a \right]$, where the functions $\chi^a(x)$ have to vanish asymptotically. 

Our construction points out that the configurations related by these emergent gauge transformations are really different physical solutions, it is only that it is difficult to operationally differentiate them. Therefore, being faithful with our construction we should not eliminate these redundancies from the path integral. This offers the first instance of a distinction between the standard Yang-Mills theory and our emergent Yang-Mills theory. Any such difference would appear in the nonperturbative regime.

\subsection{Internal and external observers}

The emergent Yang-Mills construction that we have developed leads to several interesting observations that go beyond the electrodynamics case discussed in~\cite{barcelo2016}.
When studying standard Yang-Mills theory it is usual to read that the theory does not possess meaningful local currents (see Sec. 2.6 of~\cite{jenkins2006} for a discussion of this point). We have seen that this is already a characteristic of the linear theory: there are no gauge-invariant Yang-Mills currents. In fact, this is just an instance of the Weinberg-Witten theorem \cite{weinberg1980}. However, this cannot be taken as evidence that Weinberg-Witten implies that no Yang-Mills theory can emerge from a condensed-matter-like system, the very reason behind this being the emergent nature of gauge symmetries in such a framework. If gauge symmetries are emergent, field configurations that are equivalent at low energies are not equivalent from the perspective of the high-energy theory. Alternatively, internal observers that experience only the low-energy physics cannot distinguish operationally between certain configurations, while external observers that are aware that the description used by the internal observer is limited to low energies can certainly distinguish between them (see e.g. \cite{barcelo2014a} for another example in which the distinctions between internal and external observers are discussed explicitly). Hence, it is not needed to demand the existence of certain observables (e.g. a current) that are gauge invariant as a self-consistency condition necessary for emergence. Conversely, that such a current does not exist in the low-energy description cannot be taken as an indication of the impossibility of embedding this description in an emergent framework in which the difference between configurations that are equivalent at low energies has a definite operational meaning. We hope that this clarifies that claims in the literature that the Weinberg-Witten theorem forbids the emergence of certain theories are using a too narrow notion of emergence. At most, Weinberg-Witten theorem could be taken as an indication that Lorentz and gauge symmetries must emerge simultaneously, which is indeed compatible with our discussion.

If we were only talking about interpretative choices, then the internal observer position could be argued to be superior. However, it is well known that the importance of having alternative interpretations is that they suggest different extensions when the time comes. Here we have already identified one potential physical difference between both interpretations: the need to eliminate or not Gribov copies from the path integral. This might already go beyond an interpretative issue.

\section{Discussion and conclusions}

In this work we have presented how an emergent Yang-Mills theory with an emergent gauge invariance can represent part of the dynamics of a system with more degrees of freedom and no gauge invariance a priori. Our main result is the existence of a natural coupling scheme that leads to the emergence of gauge symmetries. The logic underneath the construction is the following:

\begin{enumerate}
    \item One can start from a very complex theory, for instance from a condensed-matter-like system, with a very large but finite number of degrees of freedom.
    \item We restrict our attention to theories in which there is a low-energy regime which can be effectively described in terms of a set of weakly coupled relativistic fields $A^{a}_{\mu}$ and fermionic matter fields $\psi^i$ (examples of how this can happen can be found in~\cite{barcelo2014a} and references therein). The important point here is that some collective excitations acquire a relativistic behaviour which, in condensed-matter-like systems, is typically associated with the presence of Fermi points~\cite{volovik2008,volovik2009,geim2007}.
    \item In addition, it is typically not difficult to find situations in which the effective fields by themselves are massless (representing soundlike excitations). Later, there might be a Higgs-like mechanism giving mass to some of the fields. 
    \item Then, one can write down the most generic quadratic Lagrangian compatible with these considerations and show the existence of a symmetry under rigid transformations that provides with a conserved current. The existence of this symmetry allows us to define a nonlinear completion in which the relativistic vector fields couple to this current. For the nonlinear completion to be derivable from an action principle, the vector fields must satisfy a constraint under which the nonlinear theory develops emergent gauge symmetries. 
    
\end{enumerate}

In summary, the emergence of gauge symmetries relies only on a few simple principles: low-energy Lorentz invariance, emergence of massless vector fields describable by an action quadratic in these fields and their derivatives, and self-coupling to a conserved current associated with specific rigid symmetries. Self-consistency between these principles leads to the emergence of gauge symmetries described by semi-simple Lie-Algebras of the same dimension as the number of massless vector fields. Any theory satisfying these conditions at low energies must therefore be describable by a Yang-Mills theory, which in particular implies that these conditions suffice to guarantee the absence of classical instabilities when expanding around a classical solution to the equations of motion. Equivalently, unphysical states are removed from the nonlinear theory.

The emergence process can be understood as depending on a particular parameter, something like the temperature in a condensed matter system. Using this language, we would say that below a certain temperature this effective field theory provides a convenient description of the system; above this temperature the description could be very different and not easily related to the former. Therefore, we need not imagine the emergence procedure as something that can be described in terms of an effective field theory (with more degrees of freedom than the one presented here), and such that the Lorenz condition is nonzero in one regime of the theory but becomes zero dynamically in another regime. There are indications that dynamical mechanisms which completely suppress degrees of freedom in an effective field theory are accompanied by different pathologies.

We finish this work by recalling the structural similarity between Yang-Mills theory and general relativity. We will devote a future work to investigate whether we can extend the scheme of this paper to gravity. In fact, the motivation to analyze the electromagnetic and Yang-Mills cases comes from the gravitational case. It is indeed reasonable to assume that, if gravity is emergent, being this force special in the sense that describes the causal structure of the spacetime itself, then most surely the rest of interactions would also be emergent. In any case, we expect that the present analysis will pave the way toward the more complicated gravitational case.

\acknowledgments

G.G.M. thanks Julio Arrechea for useful comments. Financial support was provided by the Spanish Government through the projects FIS2017-
86497-C2-1-P, FIS2017-86497-C2-2-P (with FEDER contribution), FIS2016-78859-P (AEI/FEDER,UE), and by the Junta de Andaluc\'{\i}a through the project FQM219. C.B. and G.G.M. acknowledges financial support from the State Agency for Research of the Spanish MCIU through the ``Center of Excellence Severo Ochoa'' award to the Instituto de Astrof\'{\i}sica de Andaluc\'{\i}a (SEV-2017-0709). G.G.M. acknowledges financial support from IPARCOS.

\appendix

\section{SOME ALGEBRAIC CONSIDERATIONS}\label{sec:appendix}

In this appendix we show that there is no action that, after variation with respect to $A^{a\mu}$, leads to the contribution proportional to $\lambda$, namely Eq. \eqref{g-current-1}, to the current $J^{(1)a\mu}$. Let us consider a generic Lagrangian at first order
\begin{equation}
L_{(\lambda)}^{(1)}= P^{bca}_{\ \ \ \ \mu\nu\rho\sigma} \partial^\mu A^{b \nu} A^{c \rho} A^{a\sigma}.
\end{equation}
Its variational derivative  is given by
\begin{align}
 \frac{\delta L_{(\lambda)}^{(1)}}{\delta A_{a}^{\mu}}&=  P^{bca}_{\ \ \ \ \sigma\nu\rho\mu} \partial^\sigma A^{b \nu} A^{c \rho}\nonumber \\
&+P^{bac}_{\ \ \ \ \rho\nu\mu\sigma} \partial^\rho A^{b \nu} A^{c\sigma}-P^{acb}_{\ \ \ \ \nu\mu\rho\sigma} \partial^\nu \left(A^{c \rho} A^{b\sigma}\right).
\end{align}
This variational derivative can be rewritten as
\begin{align}
\frac{\delta L_{(\lambda)}^{(1)}}{\delta A_{a}^{\mu}}&=\left(P^{bca}_{\ \ \ \ \sigma\nu\rho\mu}+P^{bac}_{\ \ \ \ \sigma\nu\mu\rho} \right. \nonumber\\
&- \left. P^{abc}_{\ \ \ \ \sigma\mu\nu\rho}-P^{acb}_{\ \ \ \ \sigma\mu\rho\nu} \right) \partial^\sigma A^{b\nu}A^{c\rho}.
\end{align}
Hence, if we use the equation above for the piece of the current in Eq. \eqref{g-current-1}, we obtain the equation
\begin{equation}
    P^{bca}_{\ \ \ \ \sigma\nu\rho\mu}+P^{bac}_{\ \ \ \ \sigma\nu\mu\rho}-P^{abc}_{\ \ \ \ \sigma\mu\nu\rho}-P^{acb}_{\ \ \ \ \sigma\mu\rho\nu}\propto\gamma^{bca}\eta_{\mu\rho}\eta_{\nu\sigma}.
\end{equation}
Let us note that we have replaced $f^{bca}$ with a generic (that is, not satisfying specific antisymmetry requirements) tensor $\gamma^{bca}$.

The equation above can be further simplified using the fact that $P^{bca}_{\ \ \ \ \sigma\nu\rho\mu}$ is symmetric under the simultaneous exchange of $c\leftrightarrow a$ and $\rho\leftrightarrow \mu$. As a result, we have the simplified equation
\begin{equation}
    P^{bca}_{\ \ \ \ \sigma\nu\rho\mu}-P^{abc}_{\ \ \ \ \sigma\mu\nu\rho}\propto\gamma^{bca}\eta_{\mu\rho}\eta_{\nu\sigma}.
    \label{simplified_equation}
\end{equation}
Let us make the ansatz for $P^{bca}_{\ \ \ \ \sigma\nu\rho\mu}$ consisting in the most generic tensor linear in $\gamma^{abc}$ and quadratic in $\eta_{\mu\nu}$. The symmetries of $P^{bca}_{\ \ \ \ \sigma\nu\rho\mu}$ allow us to write the following ansatz:
\begin{widetext}
\begin{align}
    P^{bca}_{\ \ \ \ \sigma\nu\rho\mu}=
    &\eta_{\mu\rho}\eta_{\nu\sigma}\left(\alpha_1\gamma^{abc}+\alpha_2\gamma^{acb}+\alpha_3\gamma^{bac}+\alpha_3\gamma^{bca}+\alpha_2\gamma^{cab}+\alpha_1\gamma^{cba}\right)\nonumber\\
    +&\eta_{\mu\sigma}\eta_{\nu\rho}\left(\beta_1\gamma^{abc}+\beta_2\gamma^{acb}+\beta_3\gamma^{bac}+\beta_4\gamma^{bca}+\beta_5\gamma^{cab}+\beta_6\gamma^{cba}\right)\nonumber\\
    +&\eta_{\mu\nu}\eta_{\rho\sigma}\left(\beta_6\gamma^{abc}+\beta_5\gamma^{acb}+\beta_4\gamma^{bac}+\beta_3\gamma^{bca}+\beta_2\gamma^{cab}+\beta_1\gamma^{cba}\right).
\end{align}
\end{widetext}
Plugging this ansatz in Eq. \eqref{simplified_equation}, we obtain a system of algebraic equations for the coefficients $\{\alpha_i\}_{i=1}^3$ and $\{\beta_i,\}_{i=1}^6$, which turns out to be an incompatible system. Given that Noether currents are not unique as discussed above, it is necessary to check that the introduction of boundary terms given in Eq. \eqref{eq:bterms} does not allow for a solution to exist. In the presence of boundary terms, Eq. \eqref{simplified_equation} is modified to
\begin{align}\label{eq:appb}
        P^{bca}_{\ \ \ \ \sigma\nu\rho\mu}-P^{abc}_{\ \ \ \ \sigma\mu\nu\rho} &= \gamma^{bca}\eta_{\mu\rho}\eta_{\nu\sigma}\nonumber\\
        &+ B\gamma^{bca}(\eta_{\mu\nu}\eta_{\rho\sigma}-\eta_{\mu\rho}\eta_{\nu\sigma}).
\end{align}

Let us recall that, for the purposes of the discussion in this appendix, we are replacing $f^{abc}$ with a more general $\gamma^{abc}$. The analysis of the corresponding system of equations for the coefficients $\{\alpha_i\}_{i=1}^3$ and $\{\beta_i,\}_{i=1}^6$ shows also that there is no solution. 

For completeness, let us also consider the case in which $\gamma^{abc}=f^{abc}$ is totally antisymmetric, which as we have discussed in Sec. \ref{section4} appears as a necessary condition for the bootstrapping procedure to work. In this case, the most general ansatz we can make for $P^{abc}_{\ \ \ \mu \nu \rho \sigma}$ is given by Eq. \eqref{antisymmetric_ansatz}. Plugging this ansatz in Eq. \eqref{eq:appb} leads to an incompatible system for the parameters $(\beta, B )$.
Thus, we conclude that it is not possible to derive the piece of the current in Eq. \eqref{g-current-1} (plus identically conserved pieces coming from boundary terms) from a Lagrangian containing only the vector fields $A^{a\mu}$.

\bibliography{eym}

\begin{thebibliography}{43}%
\makeatletter
\providecommand \@ifxundefined [1]{%
 \@ifx{#1\undefined}
}%
\providecommand \@ifnum [1]{%
 \ifnum #1\expandafter \@firstoftwo
 \else \expandafter \@secondoftwo
 \fi
}%
\providecommand \@ifx [1]{%
 \ifx #1\expandafter \@firstoftwo
 \else \expandafter \@secondoftwo
 \fi
}%
\providecommand \natexlab [1]{#1}%
\providecommand \enquote  [1]{``#1''}%
\providecommand \bibnamefont  [1]{#1}%
\providecommand \bibfnamefont [1]{#1}%
\providecommand \citenamefont [1]{#1}%
\providecommand \href@noop [0]{\@secondoftwo}%
\providecommand \href [0]{\begingroup \@sanitize@url \@href}%
\providecommand \@href[1]{\@@startlink{#1}\@@href}%
\providecommand \@@href[1]{\endgroup#1\@@endlink}%
\providecommand \@sanitize@url [0]{\catcode `\\12\catcode `\$12\catcode
  `\&12\catcode `\#12\catcode `\^12\catcode `\_12\catcode `\%12\relax}%
\providecommand \@@startlink[1]{}%
\providecommand \@@endlink[0]{}%
\providecommand \url  [0]{\begingroup\@sanitize@url \@url }%
\providecommand \@url [1]{\endgroup\@href {#1}{\urlprefix }}%
\providecommand \urlprefix  [0]{URL }%
\providecommand \Eprint [0]{\href }%
\providecommand \doibase [0]{https://doi.org/}%
\providecommand \selectlanguage [0]{\@gobble}%
\providecommand \bibinfo  [0]{\@secondoftwo}%
\providecommand \bibfield  [0]{\@secondoftwo}%
\providecommand \translation [1]{[#1]}%
\providecommand \BibitemOpen [0]{}%
\providecommand \bibitemStop [0]{}%
\providecommand \bibitemNoStop [0]{.\EOS\space}%
\providecommand \EOS [0]{\spacefactor3000\relax}%
\providecommand \BibitemShut  [1]{\csname bibitem#1\endcsname}%
\let\auto@bib@innerbib\@empty
\bibitem [{\citenamefont {Carlip}(2001)}]{carlip2001}%
  \BibitemOpen
  \bibfield  {author} {\bibinfo {author} {\bibfnamefont {S.}~\bibnamefont
  {Carlip}},\ }\bibfield  {title} {\bibinfo {title} {{Quantum gravity: A
  Progress report}},\ }\href {https://doi.org/10.1088/0034-4885/64/8/301}
  {\bibfield  {journal} {\bibinfo  {journal} {Rept. Prog. Phys.}\ }\textbf
  {\bibinfo {volume} {64}},\ \bibinfo {pages} {885} (\bibinfo {year} {2001})},\
  \Eprint {https://arxiv.org/abs/gr-qc/0108040} {arXiv:gr-qc/0108040}
  \BibitemShut {NoStop}%
\bibitem [{\citenamefont {Thiemann}(2007)}]{thiemann2007}%
  \BibitemOpen
  \bibfield  {author} {\bibinfo {author} {\bibfnamefont {T.}~\bibnamefont
  {Thiemann}},\ }\href {https://doi.org/10.1017/CBO9780511755682} {\emph
  {\bibinfo {title} {{Modern Canonical Quantum General Relativity}}}},\
  Cambridge Monographs on Mathematical Physics\ (\bibinfo  {publisher}
  {Cambridge University Press},\ \bibinfo {year} {2007})\BibitemShut {NoStop}%
\bibitem [{\citenamefont {Polchinski}(2007{\natexlab{a}})}]{polchinski1998a}%
  \BibitemOpen
  \bibfield  {author} {\bibinfo {author} {\bibfnamefont {J.}~\bibnamefont
  {Polchinski}},\ }\href {https://doi.org/10.1017/CBO9780511816079} {\emph
  {\bibinfo {title} {{String theory. Vol. 1: An introduction to the bosonic
  string}}}},\ Cambridge Monographs on Mathematical Physics\ (\bibinfo
  {publisher} {Cambridge University Press},\ \bibinfo {year}
  {2007})\BibitemShut {NoStop}%
\bibitem [{\citenamefont {Polchinski}(2007{\natexlab{b}})}]{Polchinski1998b}%
  \BibitemOpen
  \bibfield  {author} {\bibinfo {author} {\bibfnamefont {J.}~\bibnamefont
  {Polchinski}},\ }\href {https://doi.org/10.1017/CBO9780511618123} {\emph
  {\bibinfo {title} {{String theory. Vol. 2: Superstring theory and
  beyond}}}},\ Cambridge Monographs on Mathematical Physics\ (\bibinfo
  {publisher} {Cambridge University Press},\ \bibinfo {year}
  {2007})\BibitemShut {NoStop}%
\bibitem [{\citenamefont {Barcel{\'o}}\ \emph {et~al.}(2005)\citenamefont
  {Barcel{\'o}}, \citenamefont {Liberati},\ and\ \citenamefont
  {Visser}}]{Barcelo2005}%
  \BibitemOpen
  \bibfield  {author} {\bibinfo {author} {\bibfnamefont {C.}~\bibnamefont
  {Barcel{\'o}}}, \bibinfo {author} {\bibfnamefont {S.}~\bibnamefont
  {Liberati}},\ and\ \bibinfo {author} {\bibfnamefont {M.}~\bibnamefont
  {Visser}},\ }\bibfield  {title} {\bibinfo {title} {Analogue gravity},\ }\href
  {https://doi.org/10.12942/lrr-2005-12} {\bibfield  {journal} {\bibinfo
  {journal} {Living Reviews in Relativity}\ }\textbf {\bibinfo {volume} {8}},\
  \bibinfo {pages} {12} (\bibinfo {year} {2005})}\BibitemShut {NoStop}%
\bibitem [{\citenamefont {Carlip}(2014)}]{carlip2013}%
  \BibitemOpen
  \bibfield  {author} {\bibinfo {author} {\bibfnamefont {S.}~\bibnamefont
  {Carlip}},\ }\bibfield  {title} {\bibinfo {title} {{Challenges for Emergent
  Gravity}},\ }\href {https://doi.org/10.1016/j.shpsb.2012.11.002} {\bibfield
  {journal} {\bibinfo  {journal} {Stud. Hist. Phil. Sci. B}\ }\textbf {\bibinfo
  {volume} {46}},\ \bibinfo {pages} {200} (\bibinfo {year} {2014})},\ \Eprint
  {https://arxiv.org/abs/1207.2504} {arXiv:1207.2504 [gr-qc]} \BibitemShut
  {NoStop}%
\bibitem [{\citenamefont {Witten}(2018)}]{Witten2017}%
  \BibitemOpen
  \bibfield  {author} {\bibinfo {author} {\bibfnamefont {E.}~\bibnamefont
  {Witten}},\ }\bibfield  {title} {\bibinfo {title} {{Symmetry and
  Emergence}},\ }\href {https://doi.org/10.1038/nphys4348} {\bibfield
  {journal} {\bibinfo  {journal} {Nature Phys.}\ }\textbf {\bibinfo {volume}
  {14}},\ \bibinfo {pages} {116} (\bibinfo {year} {2018})},\ \Eprint
  {https://arxiv.org/abs/1710.01791} {arXiv:1710.01791 [hep-th]} \BibitemShut
  {NoStop}%
\bibitem [{\citenamefont {Bass}(2020)}]{Bass2020}%
  \BibitemOpen
  \bibfield  {author} {\bibinfo {author} {\bibfnamefont {S.~D.}\ \bibnamefont
  {Bass}},\ }\bibfield  {title} {\bibinfo {title} {{Emergent Gauge Symmetries
  and Particle Physics}},\ }\href {https://doi.org/10.1016/j.ppnp.2020.103756}
  {\bibfield  {journal} {\bibinfo  {journal} {Prog. Part. Nucl. Phys.}\
  }\textbf {\bibinfo {volume} {113}},\ \bibinfo {pages} {103756} (\bibinfo
  {year} {2020})},\ \Eprint {https://arxiv.org/abs/2001.01705}
  {arXiv:2001.01705 [hep-ph]} \BibitemShut {NoStop}%
\bibitem [{\citenamefont {Raby}(2017)}]{raby2017}%
  \BibitemOpen
  \bibfield  {author} {\bibinfo {author} {\bibfnamefont {S.}~\bibnamefont
  {Raby}},\ }\href {https://doi.org/10.1007/978-3-319-55255-2} {\emph {\bibinfo
  {title} {{Supersymmetric Grand Unified Theories}: {From Quarks to Strings via
  SUSY GUTs}}}},\ Vol.\ \bibinfo {volume} {939}\ (\bibinfo  {publisher}
  {Springer},\ \bibinfo {year} {2017})\BibitemShut {NoStop}%
\bibitem [{\citenamefont {Lane}(1993)}]{lane1993}%
  \BibitemOpen
  \bibfield  {author} {\bibinfo {author} {\bibfnamefont {K.~D.}\ \bibnamefont
  {Lane}},\ }\bibfield  {title} {\bibinfo {title} {{An Introduction to
  technicolor}},\ }in\ \href {https://doi.org/10.1142/9789814503785_0010}
  {\emph {\bibinfo {booktitle} {{Theoretical Advanced Study Institute (TASI 93)
  in Elementary Particle Physics: The Building Blocks of Creation - From
  Microfermius to Megaparsecs}}}}\ (\bibinfo {year} {1993})\ pp.\ \bibinfo
  {pages} {381--408},\ \Eprint {https://arxiv.org/abs/hep-ph/9401324}
  {arXiv:hep-ph/9401324} \BibitemShut {NoStop}%
\bibitem [{\citenamefont {Barcel\'o}\ \emph {et~al.}(2016)\citenamefont
  {Barcel\'o}, \citenamefont {Carballo-Rubio}, \citenamefont {Di~Filippo},\
  and\ \citenamefont {Garay}}]{barcelo2016}%
  \BibitemOpen
  \bibfield  {author} {\bibinfo {author} {\bibfnamefont {C.}~\bibnamefont
  {Barcel\'o}}, \bibinfo {author} {\bibfnamefont {R.}~\bibnamefont
  {Carballo-Rubio}}, \bibinfo {author} {\bibfnamefont {F.}~\bibnamefont
  {Di~Filippo}},\ and\ \bibinfo {author} {\bibfnamefont {L.~J.}\ \bibnamefont
  {Garay}},\ }\bibfield  {title} {\bibinfo {title} {{From physical symmetries
  to emergent gauge symmetries}},\ }\href
  {https://doi.org/10.1007/JHEP10(2016)084} {\bibfield  {journal} {\bibinfo
  {journal} {JHEP}\ }\textbf {\bibinfo {volume} {10}},\ \bibinfo {pages}
  {084}},\ \Eprint {https://arxiv.org/abs/1608.07473} {arXiv:1608.07473
  [gr-qc]} \BibitemShut {NoStop}%
\bibitem [{\citenamefont {Deser}(1970)}]{deser1970}%
  \BibitemOpen
  \bibfield  {author} {\bibinfo {author} {\bibfnamefont {S.}~\bibnamefont
  {Deser}},\ }\bibfield  {title} {\bibinfo {title} {{Self-interaction and gauge
  invariance}},\ }\href {https://doi.org/10.1007/BF00759198} {\bibfield
  {journal} {\bibinfo  {journal} {Gen. Rel. Grav.}\ }\textbf {\bibinfo {volume}
  {1}},\ \bibinfo {pages} {9} (\bibinfo {year} {1970})},\ \Eprint
  {https://arxiv.org/abs/gr-qc/0411023} {arXiv:gr-qc/0411023} \BibitemShut
  {NoStop}%
\bibitem [{\citenamefont {Weinberg}\ and\ \citenamefont
  {Witten}(1980)}]{weinberg1980}%
  \BibitemOpen
  \bibfield  {author} {\bibinfo {author} {\bibfnamefont {S.}~\bibnamefont
  {Weinberg}}\ and\ \bibinfo {author} {\bibfnamefont {E.}~\bibnamefont
  {Witten}},\ }\bibfield  {title} {\bibinfo {title} {{Limits on Massless
  Particles}},\ }\href {https://doi.org/10.1016/0370-2693(80)90212-9}
  {\bibfield  {journal} {\bibinfo  {journal} {Phys. Lett. B}\ }\textbf
  {\bibinfo {volume} {96}},\ \bibinfo {pages} {59} (\bibinfo {year}
  {1980})}\BibitemShut {NoStop}%
\bibitem [{\citenamefont {Marolf}(2015)}]{marolf2015}%
  \BibitemOpen
  \bibfield  {author} {\bibinfo {author} {\bibfnamefont {D.}~\bibnamefont
  {Marolf}},\ }\bibfield  {title} {\bibinfo {title} {{Emergent Gravity Requires
  Kinematic Nonlocality}},\ }\href
  {https://doi.org/10.1103/PhysRevLett.114.031104} {\bibfield  {journal}
  {\bibinfo  {journal} {Phys. Rev. Lett.}\ }\textbf {\bibinfo {volume} {114}},\
  \bibinfo {pages} {031104} (\bibinfo {year} {2015})},\ \Eprint
  {https://arxiv.org/abs/1409.2509} {arXiv:1409.2509 [hep-th]} \BibitemShut
  {NoStop}%
\bibitem [{\citenamefont {Ba\~nados}\ and\ \citenamefont
  {Reyes}(2016)}]{banados2016}%
  \BibitemOpen
  \bibfield  {author} {\bibinfo {author} {\bibfnamefont {M.}~\bibnamefont
  {Ba\~nados}}\ and\ \bibinfo {author} {\bibfnamefont {I.~A.}\ \bibnamefont
  {Reyes}},\ }\bibfield  {title} {\bibinfo {title} {{A short review on
  Noether\textquoteright{}s theorems, gauge symmetries and boundary terms}},\
  }\href {https://doi.org/10.1142/S0218271816300214} {\bibfield  {journal}
  {\bibinfo  {journal} {Int. J. Mod. Phys. D}\ }\textbf {\bibinfo {volume}
  {25}},\ \bibinfo {pages} {1630021} (\bibinfo {year} {2016})},\ \Eprint
  {https://arxiv.org/abs/1601.03616} {arXiv:1601.03616 [hep-th]} \BibitemShut
  {NoStop}%
\bibitem [{\citenamefont {Henneaux}\ and\ \citenamefont
  {Teitelboim}(1994)}]{henneaux1992}%
  \BibitemOpen
  \bibfield  {author} {\bibinfo {author} {\bibfnamefont {M.}~\bibnamefont
  {Henneaux}}\ and\ \bibinfo {author} {\bibfnamefont {C.}~\bibnamefont
  {Teitelboim}},\ }\href {https://books.google.es/books?id=2FAuAKEKFyYC} {\emph
  {\bibinfo {title} {Quantization of Gauge Systems}}},\ Princeton paperbacks\
  (\bibinfo  {publisher} {Princeton University Press},\ \bibinfo {year}
  {1994})\BibitemShut {NoStop}%
\bibitem [{\citenamefont {Dirac}(2001)}]{dirac1964}%
  \BibitemOpen
  \bibfield  {author} {\bibinfo {author} {\bibfnamefont {P.}~\bibnamefont
  {Dirac}},\ }\href {https://books.google.es/books?id=GVwzb1rZW9kC} {\emph
  {\bibinfo {title} {Lectures on Quantum Mechanics}}},\ Belfer Graduate School
  of Science, monograph series\ (\bibinfo  {publisher} {Dover Publications},\
  \bibinfo {year} {2001})\BibitemShut {NoStop}%
\bibitem [{\citenamefont {Wipf}(1994)}]{wipf1993}%
  \BibitemOpen
  \bibfield  {author} {\bibinfo {author} {\bibfnamefont {A.~W.}\ \bibnamefont
  {Wipf}},\ }\bibfield  {title} {\bibinfo {title} {{Hamilton's formalism for
  systems with constraints}},\ }\href
  {https://doi.org/10.1007/3-540-58339-4_14} {\bibfield  {journal} {\bibinfo
  {journal} {Lect. Notes Phys.}\ }\textbf {\bibinfo {volume} {434}},\ \bibinfo
  {pages} {22} (\bibinfo {year} {1994})},\ \Eprint
  {https://arxiv.org/abs/hep-th/9312078} {arXiv:hep-th/9312078} \BibitemShut
  {NoStop}%
\bibitem [{\citenamefont {Julia}\ and\ \citenamefont
  {Silva}(1998)}]{julia1998}%
  \BibitemOpen
  \bibfield  {author} {\bibinfo {author} {\bibfnamefont {B.}~\bibnamefont
  {Julia}}\ and\ \bibinfo {author} {\bibfnamefont {S.}~\bibnamefont {Silva}},\
  }\bibfield  {title} {\bibinfo {title} {{Currents and superpotentials in
  classical gauge invariant theories. 1. Local results with applications to
  perfect fluids and general relativity}},\ }\href
  {https://doi.org/10.1088/0264-9381/15/8/006} {\bibfield  {journal} {\bibinfo
  {journal} {Class. Quant. Grav.}\ }\textbf {\bibinfo {volume} {15}},\ \bibinfo
  {pages} {2173} (\bibinfo {year} {1998})},\ \Eprint
  {https://arxiv.org/abs/gr-qc/9804029} {arXiv:gr-qc/9804029} \BibitemShut
  {NoStop}%
\bibitem [{Note1()}]{Note1}%
  \BibitemOpen
  \bibinfo {note} {It is equally possible to consider the emergence of
  additional gauge symmetries in a system in which some of the symmetries are
  already gauge; however, this makes the discussion more convoluted without
  providing additional conceptual insights.}\BibitemShut {Stop}%
\bibitem [{\citenamefont {Aldaya}\ \emph {et~al.}(2016)\citenamefont {Aldaya},
  \citenamefont {Guerrero}, \citenamefont {L\'opez-Ruiz},\ and\ \citenamefont
  {Coss\'\i{}o}}]{aldaya2016}%
  \BibitemOpen
  \bibfield  {author} {\bibinfo {author} {\bibfnamefont {V.}~\bibnamefont
  {Aldaya}}, \bibinfo {author} {\bibfnamefont {J.}~\bibnamefont {Guerrero}},
  \bibinfo {author} {\bibfnamefont {F.~F.}\ \bibnamefont {L\'opez-Ruiz}},\ and\
  \bibinfo {author} {\bibfnamefont {F.}~\bibnamefont {Coss\'\i{}o}},\
  }\bibfield  {title} {\bibinfo {title} {{$SU(2)$ particle sigma model: the
  role of contact symmetries in global quantization}},\ }\href
  {https://doi.org/10.1088/1751-8113/49/50/505201} {\bibfield  {journal}
  {\bibinfo  {journal} {J. Phys. A}\ }\textbf {\bibinfo {volume} {49}},\
  \bibinfo {pages} {505201} (\bibinfo {year} {2016})},\ \Eprint
  {https://arxiv.org/abs/1607.04058} {arXiv:1607.04058 [math-ph]} \BibitemShut
  {NoStop}%
\bibitem [{\citenamefont {Belinfante}(1940)}]{belinfante1940}%
  \BibitemOpen
  \bibfield  {author} {\bibinfo {author} {\bibfnamefont {J.}~\bibnamefont
  {Belinfante}},\ }\bibfield  {title} {\bibinfo {title} {{On the current and
  the density of the electric charge, the energy, the linear momentum and the
  angular momentum of arbitrary fields}},\ }\href@noop {} {\bibfield  {journal}
  {\bibinfo  {journal} {Physica}\ }\textbf {\bibinfo {volume} {7}},\ \bibinfo
  {pages} {449} (\bibinfo {year} {1940})}\BibitemShut {NoStop}%
\bibitem [{\citenamefont {Rosenfeld}(1940)}]{rosenfeld1940}%
  \BibitemOpen
  \bibfield  {author} {\bibinfo {author} {\bibfnamefont {L.}~\bibnamefont
  {Rosenfeld}},\ }\bibfield  {title} {\bibinfo {title} {{Sur le tenseur
  D'Impulsion-Energie}},\ }\href@noop {} {\bibfield  {journal} {\bibinfo
  {journal} {Mem. Acad. R. Belg. Sci.}\ }\textbf {\bibinfo {volume} {18}},\
  \bibinfo {pages} {1} (\bibinfo {year} {1940})}\BibitemShut {NoStop}%
\bibitem [{\citenamefont {Ortin}(2015)}]{ortin2010}%
  \BibitemOpen
  \bibfield  {author} {\bibinfo {author} {\bibfnamefont {T.}~\bibnamefont
  {Ortin}},\ }\href {https://doi.org/10.1017/CBO9781139019750} {\emph {\bibinfo
  {title} {{Gravity and Strings}}}},\ \bibinfo {edition} {2nd}\ ed.,\ Cambridge
  Monographs on Mathematical Physics\ (\bibinfo  {publisher} {Cambridge
  University Press},\ \bibinfo {year} {2015})\BibitemShut {NoStop}%
\bibitem [{\citenamefont {Padmanabhan}(2008)}]{padmanabhan2004}%
  \BibitemOpen
  \bibfield  {author} {\bibinfo {author} {\bibfnamefont {T.}~\bibnamefont
  {Padmanabhan}},\ }\bibfield  {title} {\bibinfo {title} {{From gravitons to
  gravity: Myths and reality}},\ }\href
  {https://doi.org/10.1142/S0218271808012085} {\bibfield  {journal} {\bibinfo
  {journal} {Int. J. Mod. Phys. D}\ }\textbf {\bibinfo {volume} {17}},\
  \bibinfo {pages} {367} (\bibinfo {year} {2008})},\ \Eprint
  {https://arxiv.org/abs/gr-qc/0409089} {arXiv:gr-qc/0409089} \BibitemShut
  {NoStop}%
\bibitem [{\citenamefont {Barcel\'o}\ \emph
  {et~al.}(2014{\natexlab{a}})\citenamefont {Barcel\'o}, \citenamefont
  {Carballo-Rubio},\ and\ \citenamefont {Garay}}]{barcelo2014}%
  \BibitemOpen
  \bibfield  {author} {\bibinfo {author} {\bibfnamefont {C.}~\bibnamefont
  {Barcel\'o}}, \bibinfo {author} {\bibfnamefont {R.}~\bibnamefont
  {Carballo-Rubio}},\ and\ \bibinfo {author} {\bibfnamefont {L.~J.}\
  \bibnamefont {Garay}},\ }\bibfield  {title} {\bibinfo {title} {{Unimodular
  gravity and general relativity from graviton self-interactions}},\ }\href
  {https://doi.org/10.1103/PhysRevD.89.124019} {\bibfield  {journal} {\bibinfo
  {journal} {Phys. Rev. D}\ }\textbf {\bibinfo {volume} {89}},\ \bibinfo
  {pages} {124019} (\bibinfo {year} {2014}{\natexlab{a}})},\ \Eprint
  {https://arxiv.org/abs/1401.2941} {arXiv:1401.2941 [gr-qc]} \BibitemShut
  {NoStop}%
\bibitem [{\citenamefont {Ogievetsky}\ and\ \citenamefont
  {Polubarinov}(1962)}]{Ogievetsky1962}%
  \BibitemOpen
  \bibfield  {author} {\bibinfo {author} {\bibfnamefont {V.~I.}\ \bibnamefont
  {Ogievetsky}}\ and\ \bibinfo {author} {\bibfnamefont {I.~V.}\ \bibnamefont
  {Polubarinov}},\ }\bibfield  {title} {\bibinfo {title} {{On the meaning of
  gauge invariance}},\ }\href {https://doi.org/10.1007/bf02733552} {\bibfield
  {journal} {\bibinfo  {journal} {Nuovo Cim.}\ }\textbf {\bibinfo {volume}
  {23}},\ \bibinfo {pages} {173} (\bibinfo {year} {1962})}\BibitemShut
  {NoStop}%
\bibitem [{\citenamefont {Ogievetsky}\ and\ \citenamefont
  {Polubarinov}(1966)}]{ogievetsky1965b}%
  \BibitemOpen
  \bibfield  {author} {\bibinfo {author} {\bibfnamefont {V.~I.}\ \bibnamefont
  {Ogievetsky}}\ and\ \bibinfo {author} {\bibfnamefont {I.~V.}\ \bibnamefont
  {Polubarinov}},\ }\bibfield  {title} {\bibinfo {title} {Theories of
  interacting fields with spin 1},\ }\href
  {https://doi.org/https://doi.org/10.1016/0029-5582(66)90206-9} {\bibfield
  {journal} {\bibinfo  {journal} {Nuclear Physics}\ }\textbf {\bibinfo {volume}
  {76}},\ \bibinfo {pages} {677} (\bibinfo {year} {1966})}\BibitemShut
  {NoStop}%
\bibitem [{\citenamefont {Ogievetsky}\ and\ \citenamefont
  {Polubarinov}(1965)}]{ogievetsky1965}%
  \BibitemOpen
  \bibfield  {author} {\bibinfo {author} {\bibfnamefont {V.~I.}\ \bibnamefont
  {Ogievetsky}}\ and\ \bibinfo {author} {\bibfnamefont {I.~V.}\ \bibnamefont
  {Polubarinov}},\ }\bibfield  {title} {\bibinfo {title} {{Interacting field of
  spin 2 and the Einstein equations}},\ }\href
  {https://doi.org/10.1016/0003-4916(65)90077-1} {\bibfield  {journal}
  {\bibinfo  {journal} {Annals Phys.}\ }\textbf {\bibinfo {volume} {35}},\
  \bibinfo {pages} {167} (\bibinfo {year} {1965})}\BibitemShut {NoStop}%
\bibitem [{\citenamefont {Ivanov}(2016)}]{Ivanov2016}%
  \BibitemOpen
  \bibfield  {author} {\bibinfo {author} {\bibfnamefont {E.~A.}\ \bibnamefont
  {Ivanov}},\ }\bibfield  {title} {\bibinfo {title} {{Gauge Fields, Nonlinear
  Realizations, Supersymmetry}},\ }\href
  {https://doi.org/10.1134/S1063779616040080} {\bibfield  {journal} {\bibinfo
  {journal} {Phys. Part. Nucl.}\ }\textbf {\bibinfo {volume} {47}},\ \bibinfo
  {pages} {508} (\bibinfo {year} {2016})},\ \Eprint
  {https://arxiv.org/abs/1604.01379} {arXiv:1604.01379 [hep-th]} \BibitemShut
  {NoStop}%
\bibitem [{\citenamefont {Georgi}(1999)}]{georgi1999}%
  \BibitemOpen
  \bibfield  {author} {\bibinfo {author} {\bibfnamefont {H.}~\bibnamefont
  {Georgi}},\ }\href@noop {} {\emph {\bibinfo {title} {{Lie algebras in
  particle physics}}}},\ \bibinfo {edition} {2nd}\ ed.,\ Vol.~\bibinfo {volume}
  {54}\ (\bibinfo  {publisher} {Perseus Books},\ \bibinfo {address} {Reading,
  MA},\ \bibinfo {year} {1999})\BibitemShut {NoStop}%
\bibitem [{\citenamefont {Butcher}\ \emph {et~al.}(2009)\citenamefont
  {Butcher}, \citenamefont {Hobson},\ and\ \citenamefont
  {Lasenby}}]{Butcher2009}%
  \BibitemOpen
  \bibfield  {author} {\bibinfo {author} {\bibfnamefont {L.~M.}\ \bibnamefont
  {Butcher}}, \bibinfo {author} {\bibfnamefont {M.}~\bibnamefont {Hobson}},\
  and\ \bibinfo {author} {\bibfnamefont {A.}~\bibnamefont {Lasenby}},\
  }\bibfield  {title} {\bibinfo {title} {{Bootstrapping gravity: A Consistent
  approach to energy-momentum self-coupling}},\ }\href
  {https://doi.org/10.1103/PhysRevD.80.084014} {\bibfield  {journal} {\bibinfo
  {journal} {Phys. Rev. D}\ }\textbf {\bibinfo {volume} {80}},\ \bibinfo
  {pages} {084014} (\bibinfo {year} {2009})},\ \Eprint
  {https://arxiv.org/abs/0906.0926} {arXiv:0906.0926 [gr-qc]} \BibitemShut
  {NoStop}%
\bibitem [{\citenamefont {Bhattacharjee}\ and\ \citenamefont
  {Majumdar}(2013)}]{bhattacharjee2013}%
  \BibitemOpen
  \bibfield  {author} {\bibinfo {author} {\bibfnamefont {S.}~\bibnamefont
  {Bhattacharjee}}\ and\ \bibinfo {author} {\bibfnamefont {P.}~\bibnamefont
  {Majumdar}},\ }\bibfield  {title} {\bibinfo {title} {{Gauge-free
  Coleman-Weinberg Potential}},\ }\href
  {https://doi.org/10.1140/epjc/s10052-013-2348-3} {\bibfield  {journal}
  {\bibinfo  {journal} {Eur. Phys. J. C}\ }\textbf {\bibinfo {volume} {73}},\
  \bibinfo {pages} {2348} (\bibinfo {year} {2013})},\ \Eprint
  {https://arxiv.org/abs/1302.7272} {arXiv:1302.7272 [hep-th]} \BibitemShut
  {NoStop}%
\bibitem [{\citenamefont {Basu}\ \emph {et~al.}(2018)\citenamefont {Basu},
  \citenamefont {Majumdar},\ and\ \citenamefont {Mitra}}]{basu2018}%
  \BibitemOpen
  \bibfield  {author} {\bibinfo {author} {\bibfnamefont {A.}~\bibnamefont
  {Basu}}, \bibinfo {author} {\bibfnamefont {P.}~\bibnamefont {Majumdar}},\
  and\ \bibinfo {author} {\bibfnamefont {I.}~\bibnamefont {Mitra}},\ }\bibfield
   {title} {\bibinfo {title} {Gauge-invariant matter field actions from an
  iterative noether coupling},\ }\href
  {https://doi.org/10.1103/PhysRevD.98.105018} {\bibfield  {journal} {\bibinfo
  {journal} {Phys. Rev. D}\ }\textbf {\bibinfo {volume} {98}},\ \bibinfo
  {pages} {105018} (\bibinfo {year} {2018})}\BibitemShut {NoStop}%
\bibitem [{Note2()}]{Note2}%
  \BibitemOpen
  \bibinfo {note} {By gauge orbit of a certain configuration we mean the
  subspace within field configurations that can be obtained via a gauge
  transformation, i.e., the subspace of gauge-equivalent configurations to a
  given one.}\BibitemShut {Stop}%
\bibitem [{\citenamefont {Gribov}(1978{\natexlab{a}})}]{gribov1978}%
  \BibitemOpen
  \bibfield  {author} {\bibinfo {author} {\bibfnamefont {V.~N.}\ \bibnamefont
  {Gribov}},\ }\bibfield  {title} {\bibinfo {title} {{Quantization of
  Nonabelian Gauge Theories}},\ }\href
  {https://doi.org/10.1016/0550-3213(78)90175-X} {\bibfield  {journal}
  {\bibinfo  {journal} {Nucl. Phys. B}\ }\textbf {\bibinfo {volume} {139}},\
  \bibinfo {pages} {1} (\bibinfo {year} {1978}{\natexlab{a}})}\BibitemShut
  {NoStop}%
\bibitem [{\citenamefont {Gribov}(1978{\natexlab{b}})}]{gribov1978b}%
  \BibitemOpen
  \bibfield  {author} {\bibinfo {author} {\bibfnamefont {V.~N.}\ \bibnamefont
  {Gribov}},\ }\bibfield  {title} {\bibinfo {title} {{Problem of Color
  Confinement in Nonabelian Gauge Theories}},\ }in\ \href@noop {} {\emph
  {\bibinfo {booktitle} {{5th International Seminar on High-Energy Physics
  Problems: Multiparticle Production and Limiting Fragmentation of Nuclei}}}}\
  (\bibinfo {year} {1978})\ pp.\ \bibinfo {pages} {236--260}\BibitemShut
  {NoStop}%
\bibitem [{\citenamefont {Coleman}(1985)}]{coleman1985}%
  \BibitemOpen
  \bibfield  {author} {\bibinfo {author} {\bibfnamefont {S.}~\bibnamefont
  {Coleman}},\ }\href {https://doi.org/10.1017/CBO9780511565045} {\emph
  {\bibinfo {title} {{Aspects of Symmetry}: {Selected Erice Lectures}}}}\
  (\bibinfo  {publisher} {Cambridge University Press},\ \bibinfo {address}
  {Cambridge, U.K.},\ \bibinfo {year} {1985})\BibitemShut {NoStop}%
\bibitem [{\citenamefont {Jenkins}(2006)}]{jenkins2006}%
  \BibitemOpen
  \bibfield  {author} {\bibinfo {author} {\bibfnamefont {A.}~\bibnamefont
  {Jenkins}},\ }\bibfield  {title} {\bibinfo {title} {{Topics in theoretical
  particle physics and cosmology beyond the standard model.}},\ }\href@noop {}
  {\bibfield  {journal} {\bibinfo  {journal} {{ PhD thesis, California
  Institute of Technology }}\ } (\bibinfo {year} {2006})},\ \Eprint
  {https://arxiv.org/abs/hep-th/0607239} {arXiv:hep-th/0607239} \BibitemShut
  {NoStop}%
\bibitem [{\citenamefont {Barcel\'o}\ \emph
  {et~al.}(2014{\natexlab{b}})\citenamefont {Barcel\'o}, \citenamefont
  {Carballo-Rubio}, \citenamefont {Garay},\ and\ \citenamefont
  {Jannes}}]{barcelo2014a}%
  \BibitemOpen
  \bibfield  {author} {\bibinfo {author} {\bibfnamefont {C.}~\bibnamefont
  {Barcel\'o}}, \bibinfo {author} {\bibfnamefont {R.}~\bibnamefont
  {Carballo-Rubio}}, \bibinfo {author} {\bibfnamefont {L.~J.}\ \bibnamefont
  {Garay}},\ and\ \bibinfo {author} {\bibfnamefont {G.}~\bibnamefont
  {Jannes}},\ }\bibfield  {title} {\bibinfo {title} {{Electromagnetism as an
  emergent phenomenon: a step-by-step guide}},\ }\href
  {https://doi.org/10.1088/1367-2630/16/12/123028} {\bibfield  {journal}
  {\bibinfo  {journal} {New J. Phys.}\ }\textbf {\bibinfo {volume} {16}},\
  \bibinfo {pages} {123028} (\bibinfo {year} {2014}{\natexlab{b}})},\ \Eprint
  {https://arxiv.org/abs/1407.6532} {arXiv:1407.6532 [gr-qc]} \BibitemShut
  {NoStop}%
\bibitem [{\citenamefont {Volovik}(2008)}]{volovik2008}%
  \BibitemOpen
  \bibfield  {author} {\bibinfo {author} {\bibfnamefont {G.~E.}\ \bibnamefont
  {Volovik}},\ }\bibfield  {title} {\bibinfo {title} {{Emergent physics: Fermi
  point scenario}},\ }\href {https://doi.org/10.1098/rsta.2008.0070} {\bibfield
   {journal} {\bibinfo  {journal} {Phil. Trans. Roy. Soc. Lond. A}\ }\textbf
  {\bibinfo {volume} {366}},\ \bibinfo {pages} {2935} (\bibinfo {year}
  {2008})},\ \Eprint {https://arxiv.org/abs/0801.0724} {arXiv:0801.0724
  [gr-qc]} \BibitemShut {NoStop}%
\bibitem [{\citenamefont {Volovik}(2009)}]{volovik2009}%
  \BibitemOpen
  \bibfield  {author} {\bibinfo {author} {\bibfnamefont {G.}~\bibnamefont
  {Volovik}},\ }\href {https://books.google.es/books?id=6uj76kFJOHEC} {\emph
  {\bibinfo {title} {The Universe in a Helium Droplet}}},\ International Series
  of Monographs on Physics\ (\bibinfo  {publisher} {OUP Oxford},\ \bibinfo
  {year} {2009})\BibitemShut {NoStop}%
\bibitem [{\citenamefont {Geim}\ and\ \citenamefont
  {Novoselov}(2007)}]{geim2007}%
  \BibitemOpen
  \bibfield  {author} {\bibinfo {author} {\bibfnamefont {A.}~\bibnamefont
  {Geim}}\ and\ \bibinfo {author} {\bibfnamefont {K.}~\bibnamefont
  {Novoselov}},\ }\bibfield  {title} {\bibinfo {title} {The rise of graphene},\
  }\href {https://doi.org/10.1038/nmat1849} {\bibfield  {journal} {\bibinfo
  {journal} {Nature materials}\ }\textbf {\bibinfo {volume} {6}},\ \bibinfo
  {pages} {183} (\bibinfo {year} {2007})}\BibitemShut {NoStop}%
\end{thebibliography}%

\end{document}